\documentclass[superscriptaddress, preprint, aps, prl, amsmath,amssymb]{revtex4-1}
\pdfoutput=1
\usepackage{graphicx}
\usepackage{dcolumn}
\usepackage{bm}
\usepackage{verbatim}
\usepackage{graphicx}
\usepackage{hyperref}
\usepackage{amsmath}

\newcommand{\nocontentsline}[3]{}
\newcommand{\tocless}[2]{\bgroup\let\addcontentsline=\nocontentsline#1{#2}\egroup}

\begin{document}

\title{Experimental realization and characterization of an electronic Lieb lattice}

\author{Marlou R. Slot}
\author{Thomas S. Gardenier}
\author{Peter H. Jacobse}
\affiliation{Debye Institute for Nanomaterials Science, Utrecht University, Netherlands}
\author{Guido C.P. van Miert}
\author{Sander N. Kempkes}
\affiliation{Institute for Theoretical Physics, Utrecht University, Netherlands}
\author{Stephan J.M. Zevenhuizen}
\affiliation{Debye Institute for Nanomaterials Science, Utrecht University, Netherlands}
\author{Cristiane Morais Smith}
\affiliation{Institute for Theoretical Physics, Utrecht University, Netherlands}
\author{Daniel Vanmaekelbergh}
\author{Ingmar Swart}
\affiliation{Debye Institute for Nanomaterials Science, Utrecht University, Netherlands}
\email{I.Swart@uu.nl}

\date{\today}

\maketitle
\textbf{Geometry, whether on the atomic or nanoscale, is a key factor for the electronic band structure of materials. Some specific geometries give rise to novel and potentially useful electronic bands. For example, a honeycomb lattice leads to Dirac-type bands where the charge carriers behave as massless particles \cite{CastroNeto2009}. Theoretical predictions are triggering the exploration of novel 2D geometries \cite{Weeks2010a,Guo2009,Apaja2010,Goldman2011a,Beugeling2012,Tadjine2016,VanMiert2016,Li2016,DiLiberto2016}, such as graphynes, Kagom\'{e} and the Lieb lattice. The latter is the 2D analogue of the 3D lattice exhibited by perovskites \cite{Weeks2010a}; it is a square-depleted lattice, which is characterised by a band structure featuring Dirac cones intersected by a  topological flat band. Whereas photonic and cold-atom Lieb lattices have been demonstrated \cite{Shen2010, Guzman-Silva2014,Mukherjee2015, Vicencio2015,Taie2015,Xia2016}, an electronic equivalent in 2D is difficult to realize in an existing material. Here, we report an electronic Lieb lattice formed by the surface state electrons of Cu(111) confined by an array of CO molecules positioned with a scanning tunneling microscope (STM). Using scanning tunneling microscopy, spectroscopy and wave-function mapping, we confirm the predicted characteristic electronic structure of the Lieb lattice. The experimental findings are corroborated by muffin-tin and tight-binding calculations. At higher energy, second-order electronic patterns are observed, which are equivalent to a super-Lieb lattice.}

\begin{figure}
	\centering
	\includegraphics[width=\textwidth]{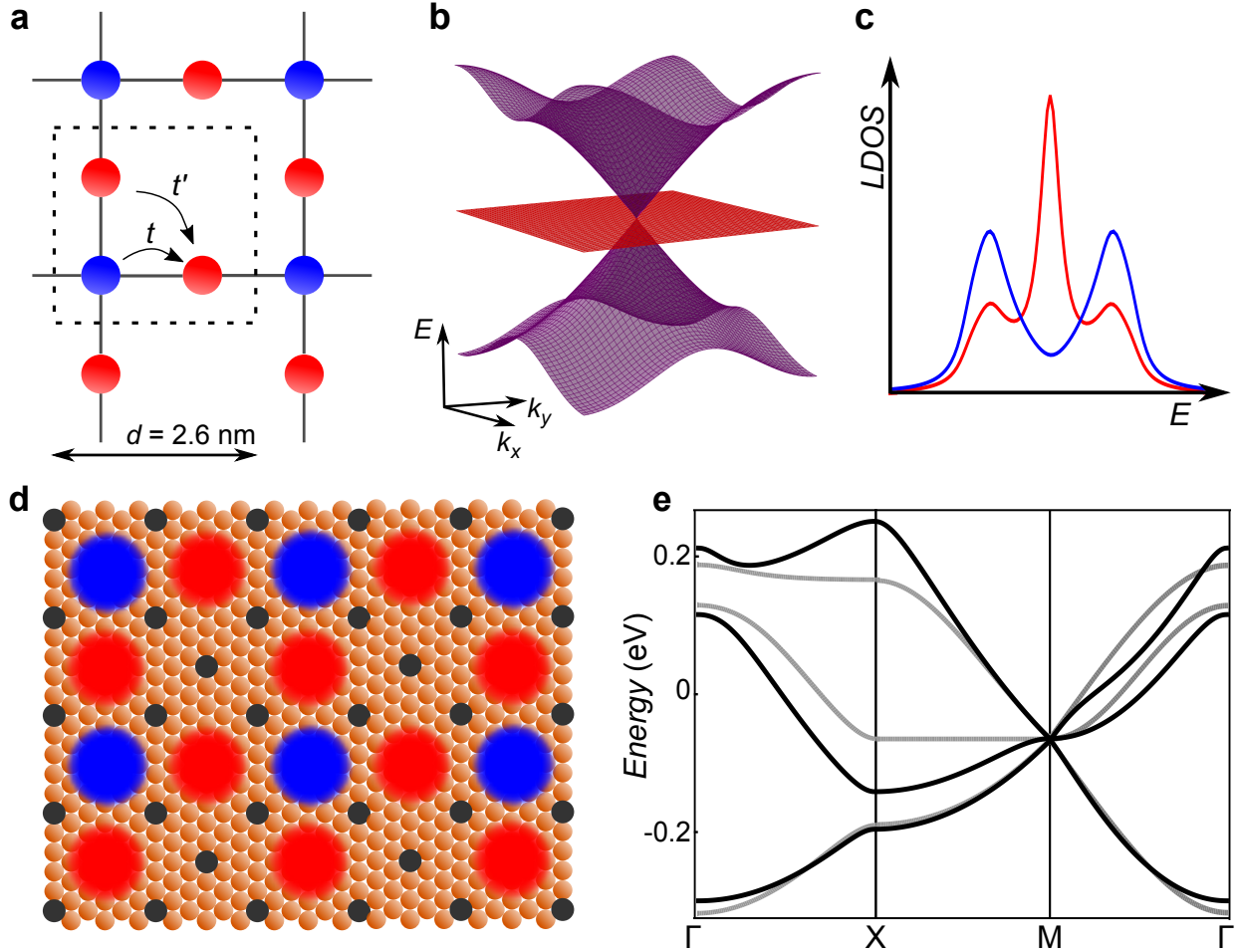}
	\caption{\textbf{Designing an electronic Lieb lattice.} \textbf{a,} Geometric structure of the Lieb lattice. The unit cell (black dashed line) contains two edge sites and one corner site, indicated in red and blue, respectively. \textbf{b,} Band structure of the Lieb lattice, only taking into account nearest-neighbor hopping. \textbf{c,} Calculated local density of states at edge (red) and corner (blue) sites. \textbf{d,} Geometric arrangement of CO molecules (black) on a Cu(111) surface to generate an electronic Lieb lattice. Red and blue circles correspond to the edge and corner sites in \textbf{a}. \textbf{e,} Band structure from muffin-tin (black) calculations along the high-symmetry lines of the Brillouin zone, overlaid with the tight-binding result using parameters that provide the best agreement with the muffin-tin simulations (gray).}
	\label{Fig1}
\end{figure}

The Lieb lattice is a square-depleted lattice, described by three sites in a square unit cell as illustrated in Fig. 1a. Two of the sites (indicated in red) are neighbored by two other sites. The third site in the unit cell (blue) has four neighbors. In the remainder of this article, these sites will be referred to as edge (red) and corner (blue) sites, respectively. This geometry results in an electronic band structure exhibiting two characteristic features: two dispersive bands, which form a Dirac cone at the $M$ point in the first Brillouin zone, and a flat band crossing the Dirac point (Fig. 1b). It is well-established that Dirac cones give rise to unusual behavior, such as effectively massless fermions. Similarly, topologically non-trivial flat bands may give rise to magnetic order \cite{Lieb1989,Costa2016}, the fractional quantum (spin) Hall and quantum anomalous Hall effect \cite{Goerbig2012,Murthy2012,Sheng2011,Zhao2012} and an enhancement of the critical temperature of superconductors \cite{Kopnin2011,Julku2016}. The electronic band structure of the Lieb lattice can be calculated from the following tight-binding Hamiltonian 
\begin{equation}
\mathcal{H} = \sum_i \epsilon_i c_i^{\dagger}c_i - t\sum_{\langle i,j\rangle}\left(c_i^{\dagger}c_j + H.c.\right) - t'\sum_{\langle\langle i,j\rangle\rangle}\left(c_i^{\dagger}c_j + H.c.\right),
\label{TBHamiltonian}
\end{equation}
where $\epsilon_i$, $t$, and $t'$ indicate the on-site energy of site $i$ and nearest- and next-nearest-neighbor hopping constants, respectively. Taking only nearest-neighbor hopping into account and using the same on-site energy for the three sites results in the band structure shown in Fig. 1b. The flat band exclusively contains electronic states which are localized on edge sites. In contrast, all sites contribute to the dispersing bands converging to the Dirac cone. Hence, the local density of states (LDOS) exhibits a characteristic spatial variation, see Fig. 1c.

Thus far, a 2-D electronic Lieb lattice has not been realized. In principle, lithography can be used to impose a Lieb pattern on a 2-D electron gas \cite{Tadjine2016}. Alternatively, a Lieb lattice can be created by assembling a molecular lattice on a substrate that features a surface state using a scanning tunneling microscope, as has been used before to prepare an artificial graphene system \cite{Gomes2012}. In the following, we will describe how atomic scale manipulation of carbon monoxide molecules on Cu(111) can be used to generate and characterize an electronic Lieb lattice.

The design of the molecular Lieb lattice is not trivial for several reasons. First, the Lieb lattice has four-fold rotational symmetry, whereas substrates that exhibit a surface state close to the Fermi energy such as Cu(111) have hexagonal symmetry. Furthermore, the CO molecules on Cu(111) act as repulsive scatterers, confining the electrons to the space between the CO molecules \cite{Gomes2012,Moon2008,Moon2009,Ropo2014}. This implies that the CO molecules should compose the anti-lattice of the electronic Lieb lattice. Our design consists of a CO square lattice, which defines the trivial anti-lattice of a square lattice, with one CO placed in the center of four CO molecules to form the anti-lattice of a depleted square lattice (\emph{cf.} Fig. 1d). This design was recently proposed independently by Qiu \emph{et al.} \cite{Qiu2016}. The size of the unit cell is chosen to be $6\sqrt{3}a_0 \times 10a_0 (\approx 2.66\,\text{nm}\times 2.56\,\text{nm})$, where $a_0 = 0.256\,$nm is the Cu(111) nearest-neighbour distance. Two factors play a critical role in the design. First, this arrangement of CO molecules provides the best approximation to the perfect four-fold symmetry of the Lieb lattice on the hexagonal Cu(111) substrate. Furthermore, the size of the unit cell determines the position of the bands of the lattice with respect to the Fermi level of the Cu(111) \cite{Gomes2012}. With the lattice constants described above, the low-energy bands of the lattice are close to the Fermi level (\emph{vide infra}).

To establish if the design described above confines the electrons in an electronic Lieb lattice, we performed calculations based on the nearly-free electron model in which the CO molecules are modelled by a muffin-tin potential. The band structure calculated using this approach is given by the black curve in Fig. 1e. These results can be reproduced well using a tight-binding model including orbital overlap and next-nearest-neighbor interactions ($t'/t = 0.6$), see the gray curve in Fig.~1e. Hence, the arrangement of CO molecules on Cu(111) shown in Fig.~1d generates an electronic Lieb lattice. The large $t'/t$ ratio shows that next-nearest-neighbor hopping in this system is important. This can be rationalized from the fact that the distance between individual CO molecules is quite large on the atomic scale. A detailed description of the correspondence between the nearly-free electron and tight-binding calculations is given in the supplementary information.

\begin{figure}
	\centering
	\includegraphics[width=\textwidth]{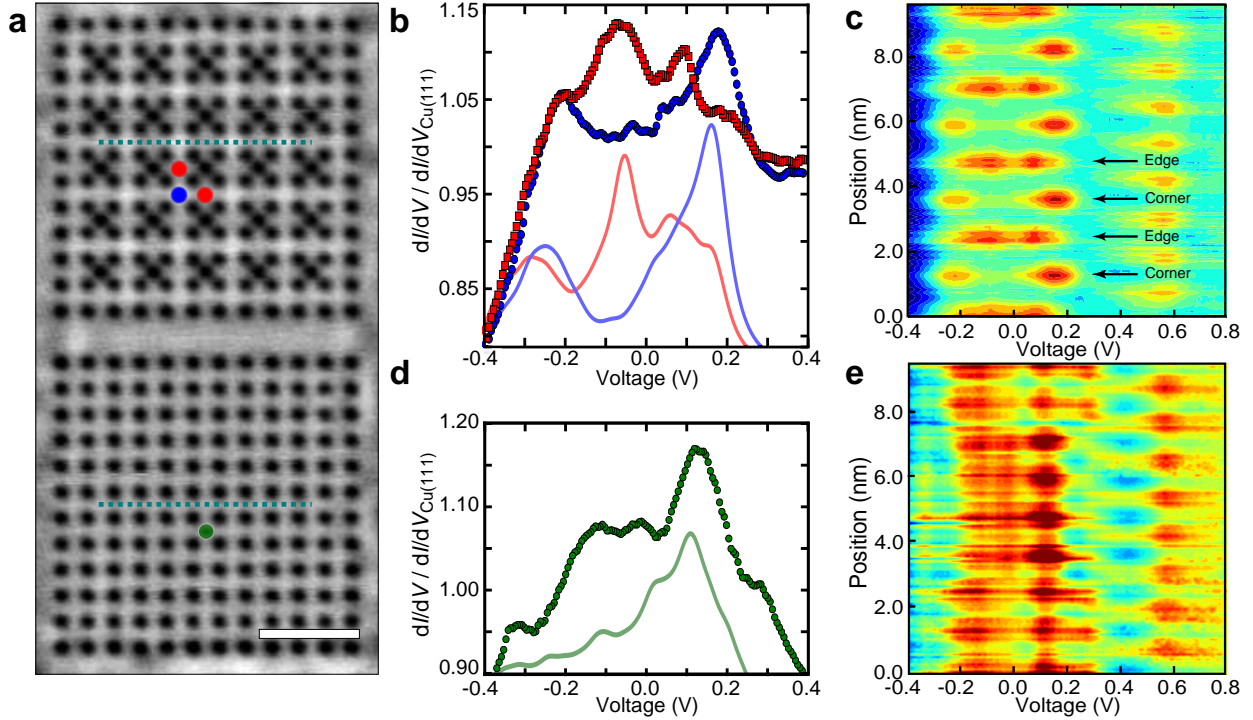}
	\caption{\textbf{Electronic structure of a Lieb lattice.} \textbf{a,} STM image of a 5x5 Lieb (top) and square (bottom) lattice. Two edge sites and one corner site of the Lieb lattice are indicated in red and blue, respectively. The green circle indicates a site of the square lattice. Imaging parameters: $V = 50\,$mV, $I = 1\,$nA. Scale bar: $5\,$nm. \textbf{b,} Normalized differential conductance spectra acquired above edge (red squares) and corner (blue circles) sites and local density of states at these sites calculated using the tight-binding method (solid lines). \textbf{c,} Contour plot of 100 spectra taken along the line indicated in \textbf{a}. The features observed in the spectra shown in \textbf{b} can be clearly recognized (see arrows). \textbf{d,e} same as \textbf{b},\textbf{c}, but for a square lattice. Note that the spectrum on the square lattice is qualitatively different from the spectra of the Lieb lattice.}
	\label{Fig2}
\end{figure}

A lattice of 5x5 unit cells was assembled in the way shown in Fig. 2a. To provide further evidence that any observed features are due to the Lieb lattice, a square lattice was created immediately next to the Lieb lattice. Differential conductance spectra were acquired above various positions of the lattice (indicated by the blue and red points in Fig.~2a). The spectra were normalized by the spectrum acquired on the clean Cu(111) surface, analogously to Ref. \cite{Gomes2012}. The resulting spectra above corner (blue) and edge sites (red) are shown in Fig.~2b. We first focus on the spectrum acquired above a corner site (blue). Two peaks are observed, one at $V = -0.20\,$V and one at $+0.18\,$V. These peaks can be assigned to the lowest- and highest-energy bands in the nearest-neighbor tight-binding model of the Lieb lattice. In between these two peaks, the LDOS reaches a minimum, which should correspond to the Dirac point. In contrast, the edge-site spectrum (red) exhibits a maximum, which is located at $V = -0.07\,$V. This peak can be assigned to the flat band. The neighboring peaks are again due to the lowest- and highest-energy bands.

In principle, a flat band should give rise to an (infinitely) narrow feature in the LDOS. In contrast, the peak at $V = -0.07\,$V observed above the edge sites is quite broad. We attribute this broadening to the influence of next-nearest-neighbor hopping, as well as due to the limited lifetime of the electrons in the surface state. 

The experimentally observed differential conductance spectra are reproduced very well when next-nearest-neighbor hopping is included in tight-binding calculations of a finite lattice (Fig. 2b). Next-nearest-neighbor hopping is essential to account for the observed asymmetry in the LDOS of the low- and high-energy bands (blue spectrum, peaks at $-0.20\,$V and $+0.18\,$V), as well as for the peak at $0.09\,$V in the edge-site spectrum. A fit of the tight-binding result to the experimental data yields $t = \left(89\pm15\right)\,$meV, which is in excellent agreement with earlier results \cite{Gomes2012}. Using this hopping parameter, we calculate the Fermi velocity of the electrons in the Dirac cones to be $v_F = \left(3.5\pm0.6\right)\cdot10^{5}\,\text{m}\cdot\text{s}^{-1}$. 

To investigate the spatial distribution of the states, we acquired differential conductance maps (\emph{vide infra}), as well 100 spectra along the line indicated in Fig. 2a. This line starts and ends at an edge site and passes four corner-sites. The resulting contour plot is shown in Fig. 2c. The peaks described above can be clearly recognized for each site, demonstrating that the LDOS features are a property of the lattice.

For comparison, a differential conductance spectrum acquired over a site in the \emph{square} lattice is shown in Fig. 2d, while a contour plot showing 125 spectra along a line are shown in Fig. 2e. The spectra along the line again demonstrate the similarity of the features for equivalent sites (Fig. 2e). Importantly, the spectra are qualitatively different from the spectra obtained over the Lieb lattice and display a good agreement with the LDOS calculated for the square lattice using the tight-binding model (using the same parameters as for the Lieb lattice), see Fig. 2d. This further demonstrates that the features observed in the differential conductance spectra shown in Fig. 2b are due to the Lieb lattice.

Figure 3 shows several experimental and simulated constant-height differential conductance maps of the two lattices. For the square lattice, all equivalent sites appear identical at all three energies. In contrast, for the Lieb lattice at $V = -0.200\,$V, both the edge and corner sites contribute significantly to the density of states. At the energy of the flat band ($V = -0.050\,$V), only the edge sites contribute. At $V = +0.150\,$V, again both corner and edge sites contribute, with the first being dominant. The simulated maps using the tight-binding model (Fig. 3d-f) and using the muffin-tin approach (Fig. 3g-i) reproduce the features observed experimentally. 

\begin{figure}
	\centering
	\includegraphics[width=\textwidth]{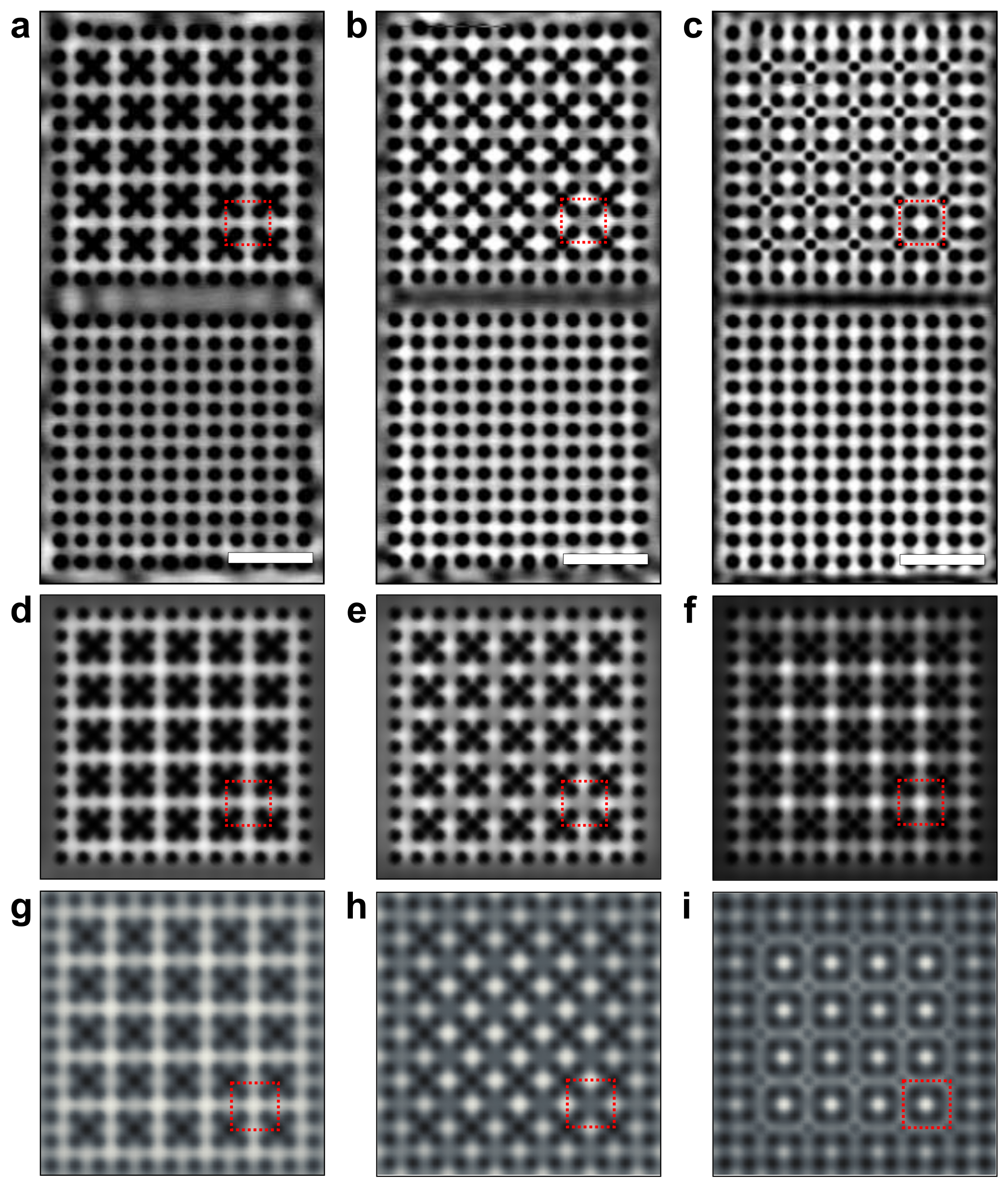}
	\caption{\textbf{Wave function mapping.} \textbf{a-c,} Differential conductance maps acquired above a Lieb and square lattice at $-0.200\,$V, $-0.050\,$V, and $+0.150\,$V, respectively. Scale bars: $5\,$nm. \textbf{d-f,} Differential conductance maps at these energies simulated using tight-binding. Black circles representing the CO molecules have been added manually. \textbf{g-i,} Same as \textbf{d-f}, but calculated using the muffin-tin model.}
	\label{Fig3}
\end{figure}

A careful inspection of the contour plots shown in Fig. 2c and 2e shows that for both the square and Lieb lattice there is structure in the spectra at higher energy (around $V = +0.600\,$V). For both lattices, these high-energy states are localized \emph{in between} different sites. To account for these states in the tight-binding calculations, additional basis functions need to be included. This can be done by adding sites in between the original sites. To first order, the simple square lattice is then described by a three-site quasi-Lieb model, with corner and edge sites having different on-site energies (Fig. 4a). Likewise, the Lieb lattice is described by a super-Lieb (Fig. 4b) geometry involving 11 sites per unit cell. Differential conductance maps of the high-energy states of the square and Lieb lattice with indicated unit cells are shown in Figs. 4c and 4d, respectively. Note that the 3 and 11 sites are required to describe the unit cells, respectively. Using this model, we again simulated differential conductance maps. This time, also the higher-energy maps are described satisfactorily. 

The peak positions with respect to the Fermi energy can be shifted to lower energies by increasing the lattice constant \cite{Gomes2012}. We make use of this effect to access even higher energy states in the square lattice. Figure 4e-h shows differential conductance maps of a square lattice with a four-times larger unit cell. For this large square lattice, the pseudo-Lieb character emerges at bias voltages as low as $-0.300\,$V and $-0.150\,$V for the bottom and flat bands, respectively. At higher bias voltages, a super-Lieb character appears, as depicted in Fig. 4g. The higher on-site energies of the 'bridging sites' results in a band gap between the lower-energy bands (which retain their square/Lieb character) and the higher-energy bands where localization is more pronounced on the bridging sites.

The ability to generate electronic lattices using CO molecules on Cu(111) allows experimental realization and characterization of lattices that have only been investigated theoretically so far. Furthermore, this system is an ideal test-bed as it allows tuning of parameters that cannot be easily varied in a real solid-state material. Combined with the ability to perform large-scale controlled manipulation of atoms and molecules automatically with atomic-scale precision \cite{Celotta2014}, we expect this approach to allow characterization of new and thus far unexplored materials.

\begin{figure}
	\centering
	\includegraphics[width=\textwidth]{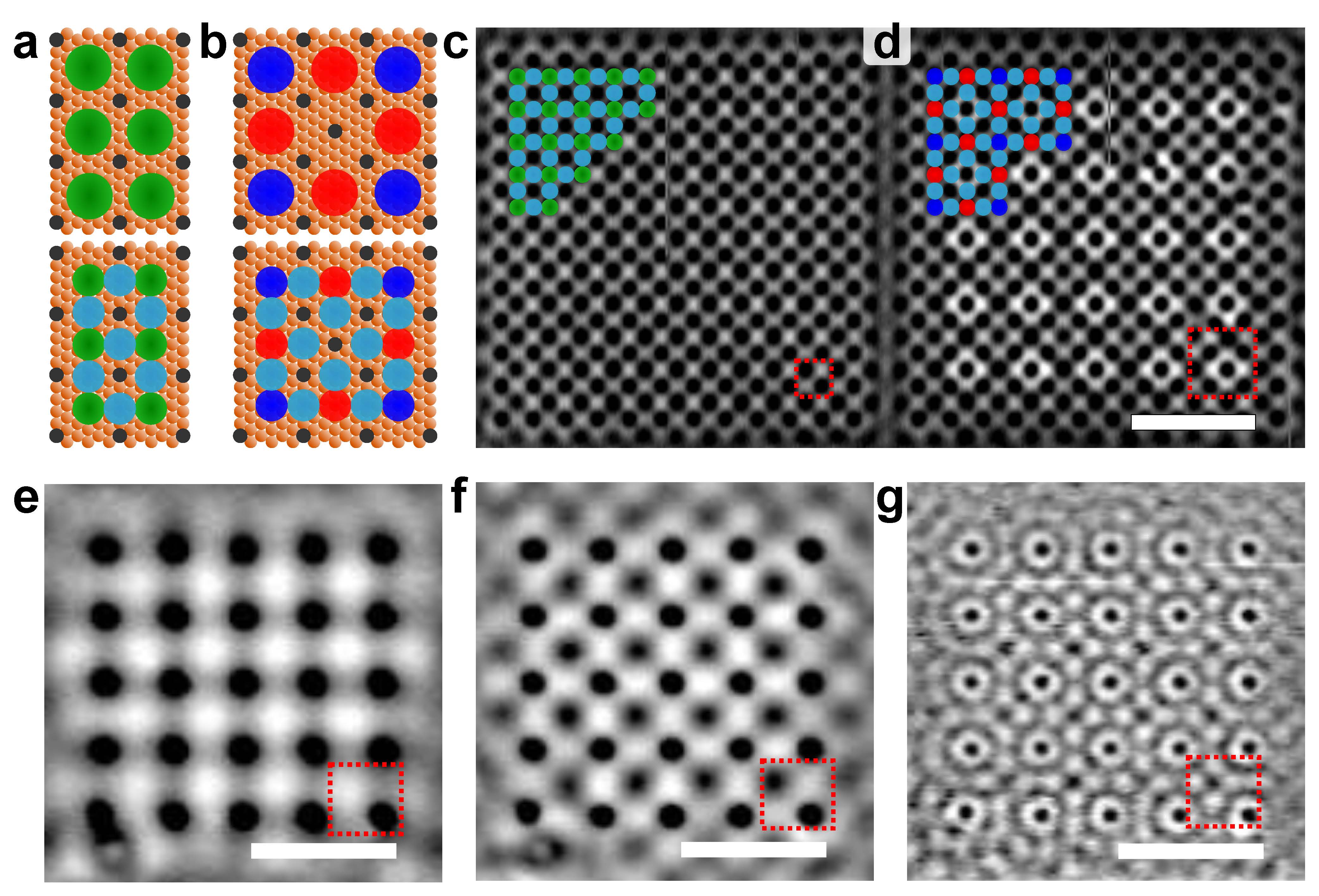}
	\caption{\textbf{Higher-order effects.} \textbf{a-b,} Sketch picture to show extra sites, resulting in a quasi-Lieb and quasi-super-Lieb lattice, respectively. \textbf{c-d,} Experimental differential conductance maps acquired at $0.550\,$V above a square and a Lieb lattice, respectively. At these energies 3 and 11 sites per unit cell are required to provide an adequate description of the wave function localization. \textbf{e-g,} Experimental differential conductance maps acquired above a square lattice at $-0.300\,$V, $-0.150\,$V, and $0.575\,$V, respectively. In each of these maps, the unit cell is indicated by a red dashed line. Note that each unit cell still only contains one CO molecule (at the bottom right of the unit cell). All scale bars denote $5\,$nm.}
	\label{Fig4}
\end{figure}

\subsection*{Methods}
\emph{STM experiments}. The experiments were performed in an ScientaOmicron LT-STM, operating at a temperature of $4.6\,$K and a pressure in the $10^{-10}\,$mbar range. Prior to the experiments, a clean Cu(111) crystal surface was prepared by several cycles of sputtering and annealing. After cooling down in the STM microscope head, CO was deposited on the surface by leaking in this gas to $P = 2\cdot 10^{-8}\,$mbar for $3\,$minutes. For all measurements a Cu coated W tip was used. Assisted by an in-house developed program, atomic manipulations were performed following previously described procedures \cite{Meyer2000,Celotta2014}. STM images were acquired in constant current mode. d$I$/d$V$ spectroscopy and mapping were performed in constant height mode using a standard lock-in amplifier modulating the sample bias with an amplitude of $10-20\,$mV rms at a frequency of $273\,$Hz.\\
The datasets generated during the current study are available from the corresponding author on reasonable request.\\
\emph{Tight-binding calculations}. Tight-binding were performed for periodic and finite-sized lattices. For dispersion and LDOS calculations, we utilized a grid of $50\times50~k$-points in the first Brillouin zone, whereas $n\times n~k$-points were used for calculating the differential conductance maps of the higher-order lattices. The used tight-binding parameters were $t'/t=0.6$ and an orbital overlap of $s=0.15$. The calculations on the experimentally realized geometry are $\Gamma$-point calculations with periodic boundary conditions, utilizing the same tight-binding parameters as the periodic lattice calculations. The local density of states was inferred in these calculations directly from the center-most sites of both lattices, using a Lorentzian energy level broadening of $\Gamma = 0.8 t$. Simulated differential conductance maps were obtained by taking again $ \Gamma = 0.8 t$ and by expanding the wave functions by normalized Gaussians of width $\sigma = 0.4 a$, where $a$ is the lattice constant of the Lieb lattice.\\
\emph{Muffin-tin calculations}. The surface state electrons of Cu(111) can be considered a two-dimensional free electron gas. The CO molecules are modeled as a disk with radius $0.3\,$nm, centered at a CO molecule, with a repulsive potential of a certain height (a parameter in the calculations). See Supplementary Information for details.

\subsection*{Acknowledgements}
Financial support from the Foundation for Fundamental Research on Matter (FOM), which is part of the Netherlands Organisation for Scientific Research (NWO), is gratefully acknowledged. We thank Joost van der Lit and Nadine van der Heijden for fruitful discussions. 

\subsection*{Author Contributions}
M.R.S., T.S.G., P.H.J. and I.S. planned the experiment, including the proposal of the design of the CO lattice. M.R.S. and T.S.G. performed the experiments and analysed the data. P.H.J. carried out the tight-binding calculations and G.C.P.M. performed the muffin-tin model calculations. S.J.M.Z. developed a program that partially automates the lattice assembly.  All authors contributed to the discussions and the manuscript.

\subsection*{Competing financial interests statement}
The authors declare no competing financial interests.


\renewcommand{\thefigure}{S\arabic{figure}}
\newpage
\section*{Supplementary Information}
\section{Muffin-tin approximation}
\subsection{Bloch Hamiltonian}
The surface state on Cu(111) can be described as a two-dimensional electron gas with an effective electron mass $m^*\approx 0.40 m_e$ \cite{Burgi2000}, where $m_e$ is the electron mass. The band bottom is located at $445\,$meV below the Fermi level $E_F$ \cite{Kroger2004}.  We account for the CO molecules by including a potential $V_{CO}$. Hence, the Hamiltonian for the system of interest reads
\begin{align*}
H&=-\frac{\hbar^2}{2m^*}\nabla^2+V_{\text{CO}}(r).
\end{align*}
Since we consider a periodic array of CO molecules, we can label the eigenstates of $H$ by a wavevector $k$ and band label $n$
\begin{align*}
H\Psi_{k,n}&=E_{k,n}\Psi_{k,n}.
\end{align*}
Bloch's theorem states that we can write the eigenstates as the product of a plane wave and a periodic function, $\Psi_{k,n}(r)=e^{ik\cdot r}u_{k,n}(r)$, where $u_{k,n}(r+a)=u_{k,n}(r)$.

We now consider the case that $V_{\text{CO}}(r)=V_0$ if $||r-r_{\text{CO}}||\leq D/2$ and zero elsewhere. This is also known as a muffin-tin potential.
The range of the potential is set to the diameter of the CO molecules as they appear on the STM maps. We find $D = 0.6\,$nm, in good agreement with DFT studies that were used for calculations on artificial graphene \cite{Ropo2014,Li2016}. For the muffin-tin potential, we choose $V_0=0.9\,$eV.

\subsection{Interactions and hybridisation with bulk states}
We should point out that we have not included the effect of interactions in our calculations. However, for the intrinsic Cu(111) surface state, it is known that electron-electron interactions combined with electron-phonon coupling leads to a broadening of approximately $20\,$meV at the band minimum of the surface state, decreasing to $0\,$meV at $E_F$ \cite{Eiguren2002}. We assume that the presence of CO molecules does not appreciably affect this broadening.\\ \\
In fact, the increased broadening of the surface states as observed in the differential conductance measurements can be attributed to the coupling of these states with the bulk, induced by the CO molecules. Due to the increased unit cell, the bulk and surface states overlap in both energy and momentum. It follows from Fermi's golden rule that the linewidth $\Gamma$ is given by
\begin{align}
\Gamma&=2|\langle b|V_{CO}|s\rangle|^2 \rho_b,
\end{align}
where $|s\rangle$ ($|b\rangle$) is a surface (bulk) state and $\rho_b$ the bulk density of states into which the surface state $|s\rangle$ can scatter. It was shown in Ref.~\cite{Gomes2012} that this can be approximated by
\begin{align}
\Gamma&=2\frac{V_0}{W}\left(\frac{a}{a_{\text{Cu}}}\right)^2,
\end{align}
where $W$ is the copper bandwidth and $a$ ($a_{\text{Cu}}$), the superlattice (copper) lattice constant. This approximation is valid for a superlattice with one CO molecule per unit cell. Using $V_0\approx W\approx 1$eV, we find that for the large ($a = 2.66\,$nm) square lattice $\Gamma\approx 80\,$meV, and for the small ($a = 1.33\,$nm) square lattice $\Gamma\approx 20\,$meV. The factor $\Gamma_{\text{large}} / \Gamma_{\text{small}} \approx 4$ can be understood as follows: the total potential for the small square lattice contains $4$ times as many scatterers, which leads to a factor of $4^2=16$. On the other hand, for the large square lattice there are $4$ times as many states to decay in due to the increased unit cell. Hence, $\rho_b$ for the small square lattice is four times as small. By combining the two effects, we thus find an overall factor of four. We assume that the large square lattice and the Lieb lattice have a comparable linewidth, i.e. $\Gamma\approx 80\,$meV.

\subsection{Results for the Lieb lattice}
\begin{figure}[!h]%
    \centering
    \includegraphics[width=0.7\textwidth]{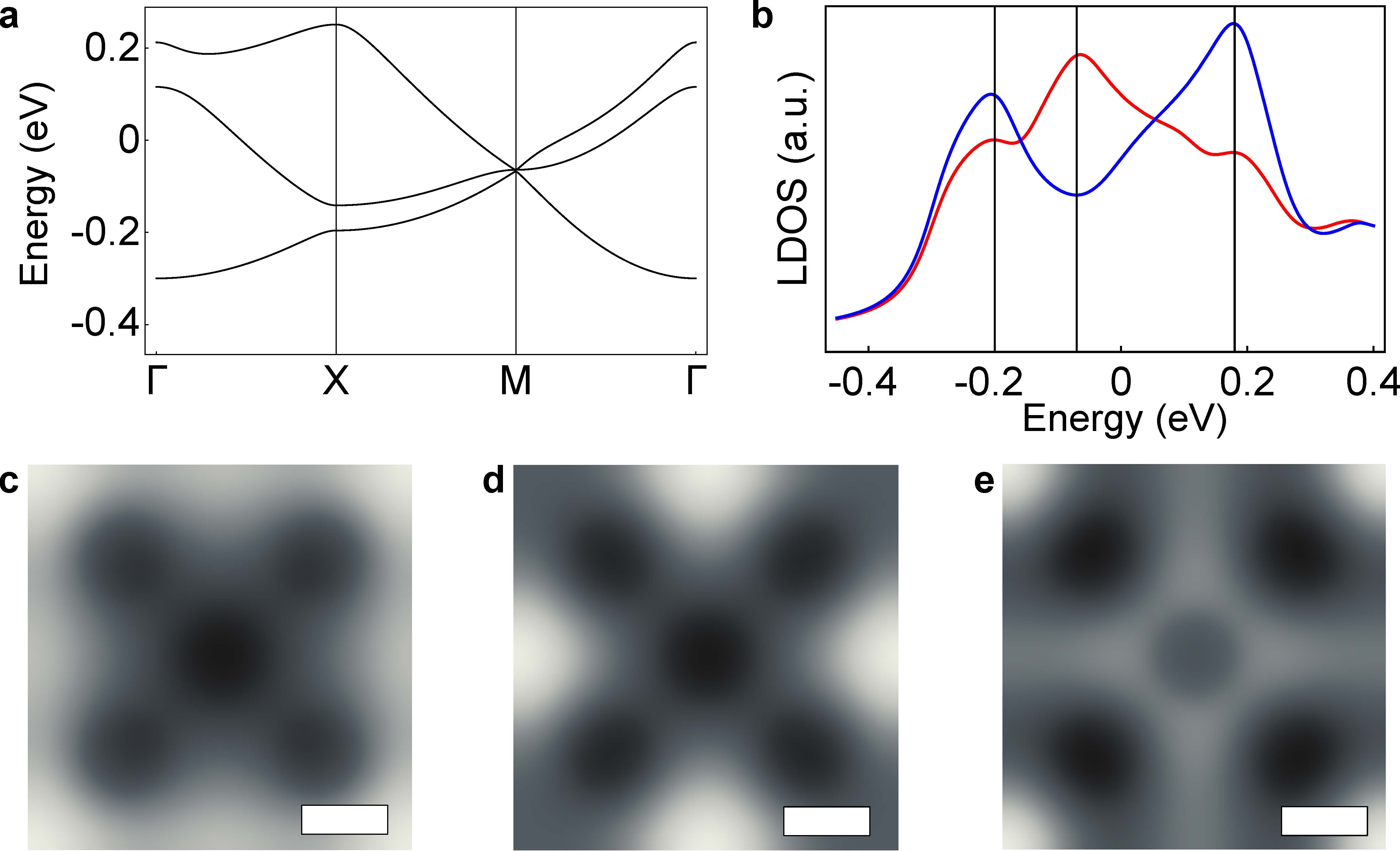}
		\caption{\textbf{Lieb lattice.} \textbf{a,} Band structure along high-symmetry lines based on a muffin-tin approximation with $V_0=0.9\,$eV, $D=0.6\,$nm and $a=2.66\,$nm. \textbf{b,} Corresponding LDOS for an edge (red) and corner (blue) site. The vertical lines correspond to the first three Van Hove singularities. We included a broadening of $80\,$meV. \textbf{c,d,e,} Simulated maps at $-0.200\,$eV, $-0.070\,$eV, and $0.180\,$eV, respectively, corresponding to the Van Hove singularities. The scale bars are $0.6\,$nm.}%
    \label{fig:1}%
\end{figure}

Calculations are performed for a geometry including  $5$ CO molecules per unit cell of $2.66\,\text{nm} \times 2.66\,\text{nm}$ using the described parameters in the muffin-tin approximation. The band structure is shown in Fig.~\ref{fig:1}a. We find that the three lowest bands are degenerate at the $M$ point. Moreover, this degeneracy shows up in the LDOS as a peak for the edge sites and a V-shaped DOS for the corner sites, see Fig.~\ref{fig:1}b. These features are also observed in the simulated STM maps, as shown in Fig.~\ref{fig:1}c-e.

\subsection{Finite-size effects}
\begin{figure}[!ht]%
    \centering
    {{\includegraphics[width=\textwidth]{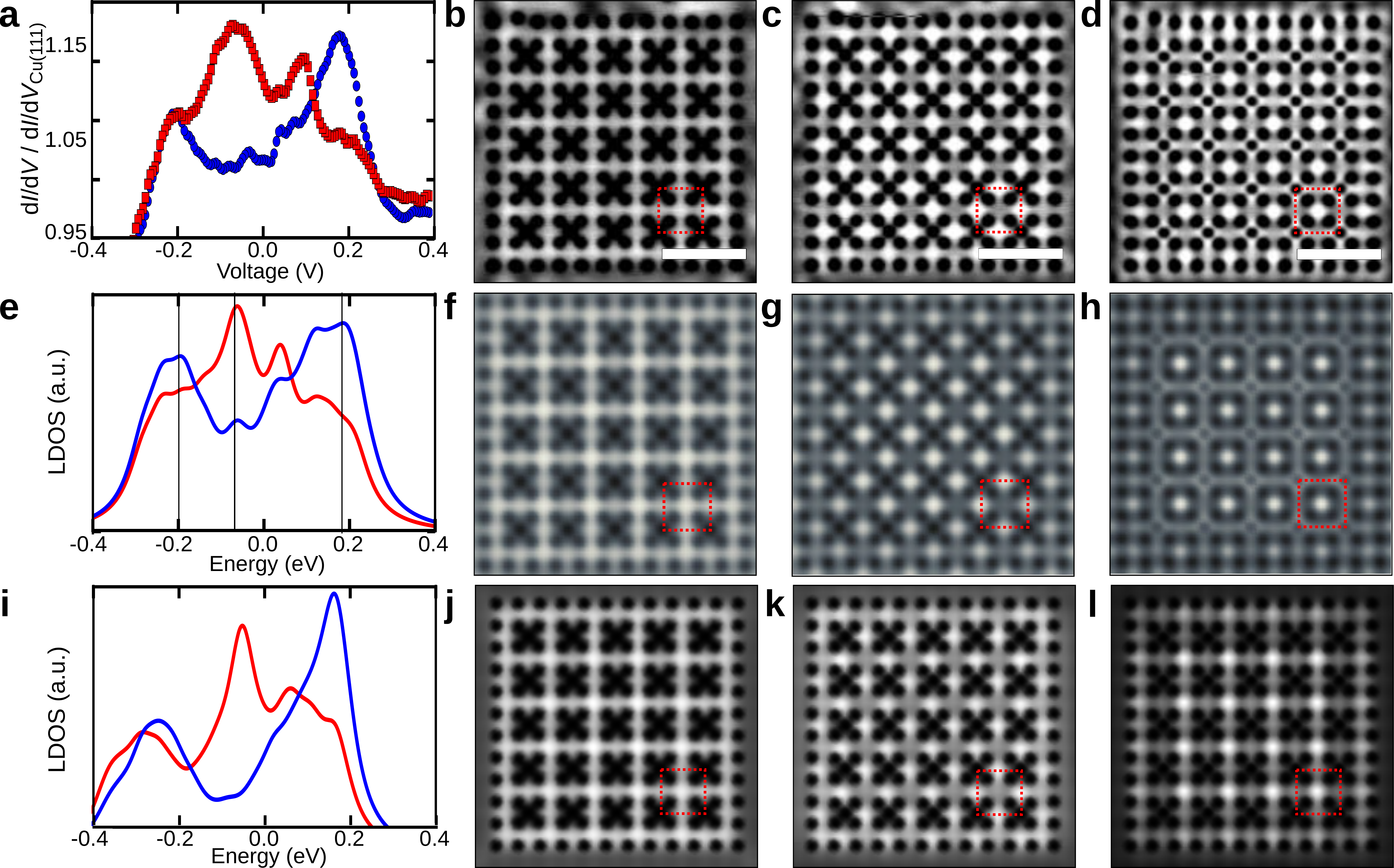} }}%
    \caption{\textbf{5~x~5 Lieb lattice.} \textbf{a,e,i,} LDOS for the edge (red) and corner (blue) sites for the finite system in the STS experiment, the muffin-tin approximation, and the tight-binding calculations, respectively. We included a broadening of $80\,$meV ($71\,$meV) in the muffin-tin (tight-binding) simulations. \textbf{b-d,f-h,j-l,} Maps for the finite system at approximately $-0.20\,$eV, $-0.07\,$eV, and $0.18\,$eV, the energies of the Van Hove singularities of the periodic system, for the experiment (\textbf{b-d}, scale bar: $5\,$nm), muffin-tin (\textbf{f-h}), and tight-binding calculations (\textbf{j-l}). The red dashed square indicates the unit cell of the Lieb lattice, centered at a corner site. }
		\label{fig:3}
\end{figure}

For the artificial graphene system, it has been shown that the DOS of the finite system converges gradually to the DOS of the fully periodic system while increasing the number of unit cells \cite{Aichinger2014,Kylanpaa2015}. To study the role of finite-size effects in the experimentally realized Lieb lattice, we have solved the Schr\"odinger equation for the finite system with the same Lieb lattice geometry as in the experiment, \emph{i.e.} a Lieb lattice with $5~\times~5$ unit cells surrounded by a 2DEG. We employed periodic boundary conditions for this entire system. The resulting LDOS for the edge and corner sites is displayed in Fig.~\ref{fig:3}e and the 
simulated LDOS maps at the energies of the lowest-energy, flat and higher-energy band are shown in Fig.~\ref{fig:3}f-h. Both the LDOS and the simulated maps are in excellent agreement with the experimental data and tight-binding simulations shown in Figs.~\ref{fig:3}a-d and~\ref{fig:3}i-l, respectively. In particular, the edge-spectrum peak between the peaks assigned to the flat band ($-0.07\,$eV) and the highest-energy band ($0.18\,$eV) was reproduced.

\subsection{Results for the large square lattice}
\begin{figure}[!ht]%
    \centering
    {{\includegraphics[width=0.7\textwidth]{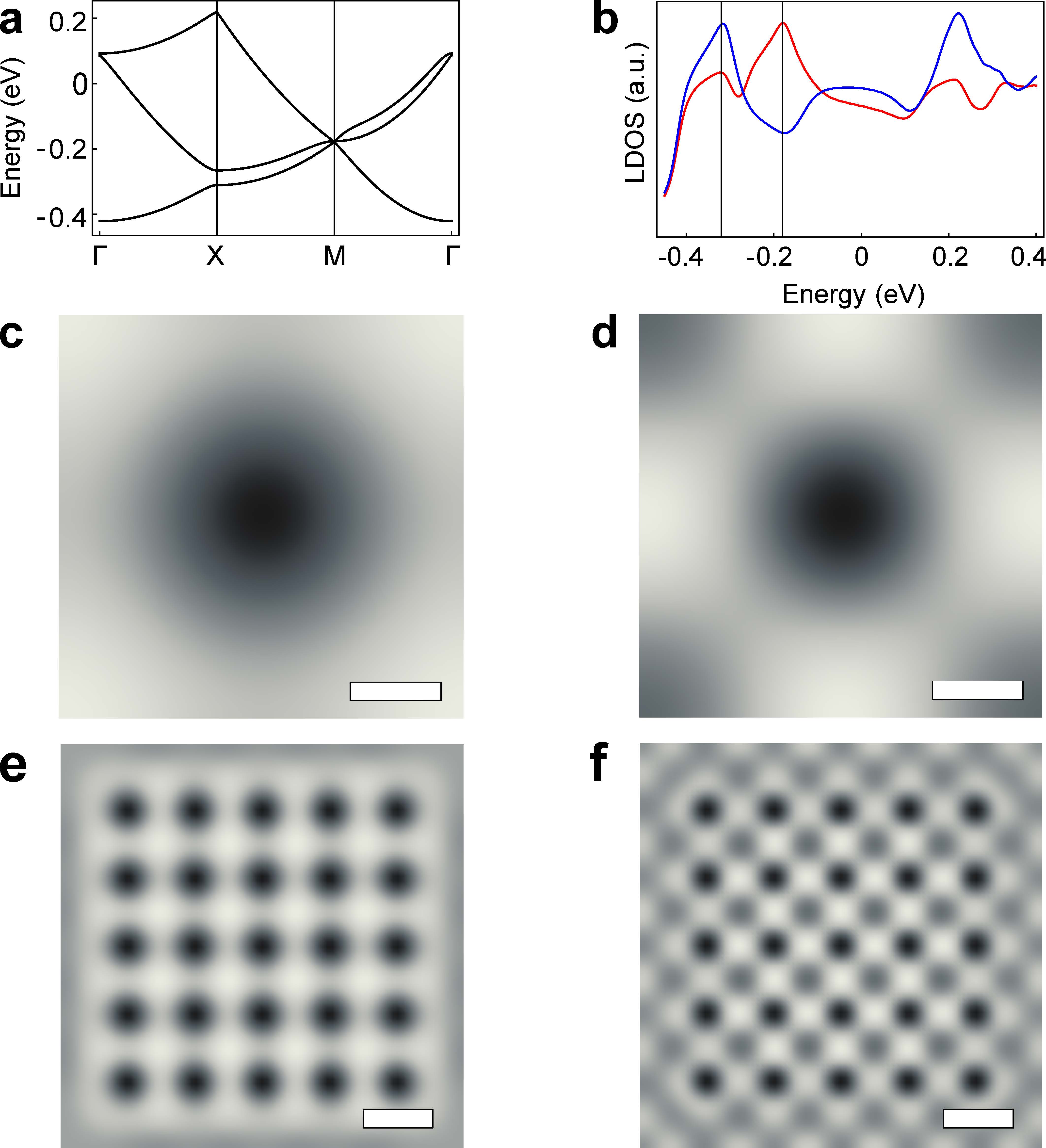} }}%
    \caption{\textbf{Large square lattice.} \textbf{a,} Band structure of the large square lattice along high-symmetry lines. \textbf{b,} LDOS for pseudo-edge (red) and pseudo-corner (blue) sites. The two vertical lines correspond to the first two Van Hove singularities. We included a broadening of $40\,$meV. \textbf{c,d} \textbf{(e,f)} Simulated maps at $-0.320\,$eV and $-0.180\,$eV for the infinite (finite) system, including a broadening of $80\,$meV. Scale bar: $0.6\,$nm ($2.66\,$nm).}
		\label{fig:5}
\end{figure}

We have also calculated the band structure for the large square lattice using the same parameters, see Fig.~\ref{fig:5}a. Note that here we also find a triple degeneracy at the $M$ point. In addition, the LDOS at the pseudo-edge and pseudo-corner sites shows that the electrons are localized on the pseudo-edge sites at the energy corresponding to the $M$ point (\emph{cf.} Fig.~\ref{fig:5}b). The simulated maps in Fig.~\ref{fig:5}c-d show this localization correspondingly. For the finite square lattice, which comprises 5~x~5 unit cells, the simulated maps are shown in Fig.~\ref{fig:5}e-f and agree with the experimental data very well (\emph{cf.} Fig.~4e-f of the main text).

\clearpage
\newpage
\section{Tight-binding model}

\subsection{Results from the tight-binding model}

\begin{figure}[!ht]
  \centering
  \includegraphics[width=.8\textwidth]{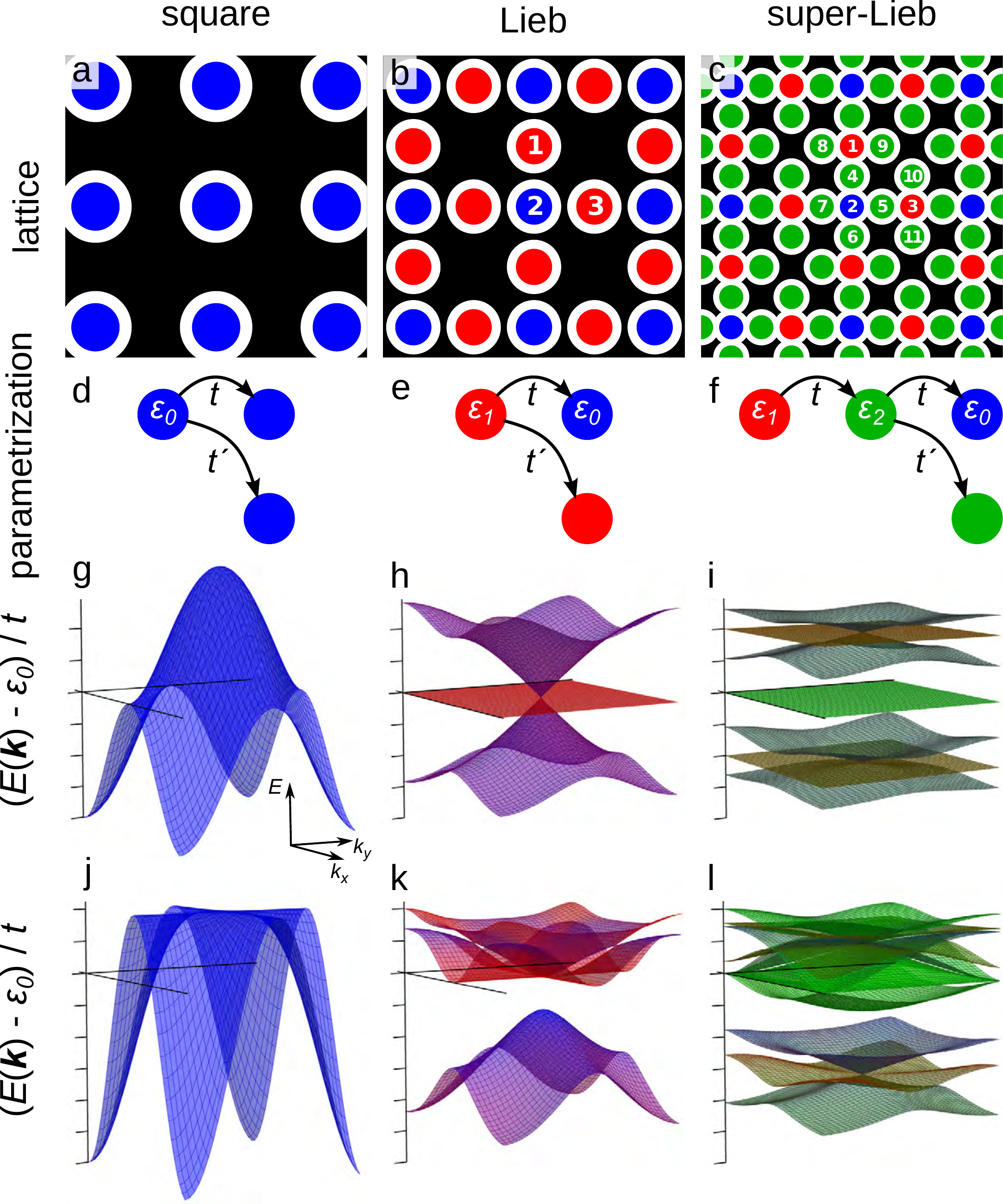}
  \caption{\textbf{Tight-binding model of the square, Lieb and super-Lieb lattice.} \textbf{a-c,} Real-space model of the respective lattices.  \textbf{d-f,} Tight-binding parametrizations.  \textbf{g-i,} Band dispersion diagrams for the respective lattices for the case $\epsilon_{0} = \epsilon_{1} = \epsilon_{2}, t' = 0$.  \textbf{j-l,} Band dispersion diagrams for the respective lattices for the case $t' = t/2$, with $\epsilon_1 = \epsilon_0 - t$ for panel~\textbf{k} and with $\epsilon_2 = \epsilon_0 - t = \epsilon_1 - t$. The local hue of the bands represents the degree of localization of the corresponding crystal orbital on the 'corner sites' (blue), 'edge sites' (red) and 'bridging sites' (green).}
	\label{squareLiebsuperLieb}
\end{figure}

Consider the tight-binding Hamiltonian
\[
\mathcal{H} = \sum_i \epsilon_i a_i^{\dagger}a_i - t\sum_{\langle i,j\rangle}\left(a_i^{\dagger}a_j + H.c.\right) - t'\sum_{\langle\langle i,j\rangle\rangle}\left(a_i^{\dagger}a_j + H.c.\right),
\]
operating on the following Bloch wave function,

\[| \psi \rangle = \sum_{m,n}^{unit\ cells} \mathbf{c}\cdot \vec{| \chi \rangle} e^{i \mathbf{k} \cdot \mathbf{r}}= \sum_{m,n}^{unit\ cells} \sum_{i}^{sites} c_i | i \rangle e^{i \mathbf{k} \cdot \mathbf{r}},\]
where $\vec{| \chi \rangle} = (| 1 \rangle, |2 \rangle, ...)$ is a basis of site-localized orbitals, $\mathbf{c}=(c_1,c_2,...)$ is the vector of expansion coefficients and $t$ and $t'$ are the respective hopping parameters for nearest-neighbor and next-nearest-neighbor hopping. Here, the nearest neighbors are defined as the horizontal and vertical neighbors on the lattices, as shown in Fig.~\ref{squareLiebsuperLieb}. The next-nearest neighbors are the diagonal neighbors. The tight-binding Hamiltonian is operated on the Bloch wave function of the square lattice, Lieb lattice and super-Lieb lattice to bring the Schr\"odinger equation $\hat{H}|\psi \rangle = E|\psi \rangle$ into the matrix form $\mathbf{H}\mathbf{c} = E \mathbf{c}$. \\ \\
For the square lattice, we obtain the following (scalar, or 1 by 1 matrix) Hamiltonian

\[\mathbf{H_{square}} = \epsilon_0 - 2t \left[\cos \left(k_x a\right) + \cos \left(k_y a\right)\right] - 4t \cos \left(k_x a\right) \cos \left(k_y a\right),\]
whereas for the Lieb lattice, the (matrix) Hamiltonian reads

\[
\mathbf{H_{Lieb}}=\left(
\begin{array}{ccc}
 \epsilon_1 & -t \gamma_y & -t' \gamma^*_x \gamma_y \\
 -t \gamma^*_y & \epsilon_0 & -t \gamma^*_x \\
 -t' \gamma_x \gamma^*_y & -t \gamma_x & \epsilon_1 \\
\end{array}
\right),
\]
where $\gamma^{(*)}_{x,y} = 1+e^{(-)i k_{x,y} a}$. Here, we have assumed that the lattice is square, such that sites $1$ and $3$ are equivalent from a symmetry perspective, but not necessarily with the same on-site energy as the site $2$. For the super-Lieb lattice, we obtain the following (matrix) Hamiltonian
\[\mathbf{H_{super-Lieb}} = \left(
\begin{array}{ccccccccccc}
 \epsilon_1  & 0 & 0 & -t & 0 & -t \xi^{*}_{y} & 0 & -t & -t & 0 & 0 \\
 0 & \epsilon_0 & 0 & -t & -t & -t & -t & 0 & 0 & 0 & 0 \\
 0 & 0 & \epsilon_1 & 0 & -t & 0 & -t \xi^{*}_{x} & 0 & 0 & -t & -t \\
 -t & -t & 0 & \epsilon_2 & -t' & 0 & -t' & -t' & -t' & 0 & 0 \\
 0 & -t & -t & -t' & \epsilon_2 & -t' & 0 & 0 & 0 & -t' & -t' \\
 -t \xi_{y} & -t & 0 & 0 & -t' & \epsilon_2 & -t' & -t' \xi_{y} & -t' \xi_{y} & 0 & 0 \\
 0 & -t & -t \xi_{x} & -t' & 0 & -t' & \epsilon_2 & 0 & 0 & -t' \xi_{x} & -t' \xi_{x} \\
 -t & 0 & 0 & -t' & 0 & -t' \xi^{*}_{y} & 0 & \epsilon_2 & 0 & -t' \xi_{x} & -t' \xi_{x} \xi^{*}_{y} \\
 -t & 0 & 0 & -t' & 0 & -t' \xi^{*}_{y} & 0 & 0 & \epsilon_2 & -t' & -t' \xi^{*}_{y} \\
 0 & 0 & -t & 0 & -t' & 0 & -t' \xi^{*}_{x} & -t' \xi^{*}_{x} & -t' & \epsilon_2 & 0 \\
 0 & 0 & -t & 0 & -t' & 0 & -t' \xi^{*}_{x} & -t' \xi^{*}_{x} \xi_{y} & -t' \xi_{y} & 0 & \epsilon_2 \\
\end{array}
\right),\]
where $\xi^{(*)}_{x,y} = e^{(-)i k_{x,y} a}$. 
Here the first three rows and columns describe the Lieb sites (indicated by red and blue in Fig.~\ref{squareLiebsuperLieb}), which are coupled to the bridging sites (green in Fig.~\ref{squareLiebsuperLieb}) with hopping parameter $t$. The bottom-right block of 8 by 8 elements contains hopping between bridging sites with next-nearest-neighbor hopping $t'$.
\\ \\ The tight-binding band structures were computed by diagonalizing the corresponding matrix Hamiltonian, and are shown in Fig.~\ref{squareLiebsuperLieb}. For each band and each $k$-value, we have furthermore considered the degree of localization of the crystal orbital on the symmetry-inequivalent sites, which are called the 'corner sites' (blue), 'edge sites' (red), and 'bridging sites' (green). The local hue of the bands represents the extent to which the wave function is localized on one of these families of sites. In Fig.~\ref{squareLiebsuperLieb}g-i, the dispersions are shown for an idealized parametrization in which all on-site energies are equal and next-nearest-neighbor hopping is set to zero. For the Lieb lattice, the top and bottom bands touch at the Dirac cone, which intersects a flat band at the Fermi level. As can be inferred from the red color of the flat band, all electronic states within the flat band are localized exclusively on the bridging sites, whereas the dispersive bands are hybrids of the corner and bridging sites. For the super-Lieb lattice, the dispersion can be described as two `Lieb-like' band structures separated by a band gap. In the middle of the gap, a set of flat bands arise, which are combinations of states localized on the bridging sites. In the case of half-filling, the square lattice may be classified as being metallic, the Lieb lattice as semimetallic and the super-Lieb as a semiconductor. \\ \\
Fig.~\ref{squareLiebsuperLieb}j-k shows the same dispersions for the more general case, where the symmetry-inequivalent sites have different on-site energies and next-nearest-neighbor hopping is significant. It has been shown that the Dirac cone in the Lieb dispersion is resilient towards changes in the next-nearest-neighbor hopping, but splits when the on-site energies of the corner and edge sites are different. On the other hand, the flat band remains unaffected by asymmetry in the on-site energies, whereas it becomes dispersive when next-nearest-neighbor hopping is introduced. Both effects can be shown to be true easily when analyzing the Hamiltonian matrix in more depth, which we will show later. The emerging dispersion of the flat band has a characteristic pattern in which it is curved up around the $\mathbf{\Gamma}$-point but remains flat on the lines connecting $\mathbf{M}$ and $\mathbf{X}$. Note that this pattern also emerges in the flat bands of the Lieb-like bands in the super-Lieb dispersion. Here, we do not assume any direct next-nearest neighbor hopping between the edge (red) sites, but the effect is introduced indirectly through the hopping of the bridging (green) sites. We note that no discernible band gap has been observed in the spectroscopy experiments on the Lieb lattices, which suggests $\Delta \epsilon \approx 0$ for the lattice set up by CO molecules, at least within the experimental broadening. As a result, our tight-binding fit could be limited to finding the best values of $t'/t$. To account for the wave functions not being entirely orthogonal, a small nearest-neighbor overlap of $s=0.15$ is included to obtain a good fit to the experimental data. \\ \\
An important point to notice is that the bottom bands of the super-Lieb lattice are reminiscent of the total dispersion of the Lieb lattice, and have a similar localization onto the corner and edge sites. These bands may therefore be thought to have a significant Lieb-like character, which becomes even stronger when the bridging sites are higher in energy than the corner and edge sites. Similarly, an effective square dispersion emerges in the lowest band of the Lieb dispersion when shifting the edge sites up in energy. The changing hue of the bottom band in Fig.~\ref{squareLiebsuperLieb}k confirms that the wave function becomes more localized on the corner sites, which really form a square lattice. This justifies our extension of the square lattice to the Lieb lattice and the Lieb lattice to the super-Lieb lattice when analyzing higher-order energy effects. It should be kept in mind that the tight-binding method only yields as many bands as the number of sites in the model. Hence, in order to look at higher-energy effects, it makes sense to add interstitial sites with higher on-site energy. The addition of edge or bridging sites allows one to look at the higher-energy bands of the lattice, whereas we have shown that the low-energy features are still described by the original bands. Effectively, the low-energy bands are now contained in the model as pseudo-square bands of the Lieb lattice and pseudo-Lieb bands of the super-Lieb lattice. Additional justification for the addition of interstitial sites is given below.

\subsection{Tight-binding as a description of the kinetic-energy landscape}

The tight-binding model can be thought of as a discretization of the Schr\"odinger equation into a set of sites that are connected through parametrized interaction integrals. Given a spatially varying potential-energy landscape, the most logical conversion into a tight-binding model is therefore obtained by assigning sites to the valleys, where the low-energy electrons are mostly localized. As such, the tight-binding model may intuitively carry the connotation of a discretization of the potential-energy landscape, where the hopping parameter fulfills the role of describing the potential energy ``hill'' that electrons have to climb over to hop from one site to the next.

As inferred from density functional theory, the potential-energy landscape that is the background of the surface-state electrons is rather flat. Therefore, the intuitive approach of assigning tight-binding sites to valleys in the potential-energy landscape is invalid. Nevertheless, the results from the muffin-tin potential landscape show that a Lieb-like band structure can still be obtained. The band structure may therefore be described as arising from a ``kinetic-energy landscape'', that results from local confinement of the electronic states.

To strengthen the idea of the ``kinetic-energy landscape'', we calculated the band structure for two toy models of electronic waveguides. In both models, the electronic states are perturbed from free-electron waves through application of a harmonic modulation. In the first waveguide, this modulation was manifested in the background potential $V(x)$, whereas in the second model, the width of the waveguide $w(x)$ was modulated while the background potential was set to zero. These two models allow to compare the effects of modulating the potential-energy landscape to modulating the kinetic-energy landscape.

Waveguide 1 is defined by the potential
\[V(x,y) = \left\lbrace \begin{array}{c}
V_0 \cos x, \qquad |y| \leq 3/2 \\
\infty, \qquad |y| > 3/2 \\
\end{array} \right\rbrace.\]
Here, the wave functions were expanded from
\[| \psi \rangle(q,n,k) = \left\lbrace \begin{array}{c}
\frac{1}{2} e^{i \left(\frac{k}{2\pi} + q \right)x} \left(e^{i \frac{n \pi y}{3}} + (-1)^{n+1} e^{-i \frac{n \pi y}{3}}\right), \qquad |y| \leq 3/2 \\
0, \qquad |y| > 3/2 \\
\end{array} \right\rbrace.\]
Wave guide 2 is defined by the potential
\[V(x,y) = \left\lbrace \begin{array}{c}
0, \qquad |y| \leq (3 + \cos x)/2 \\
\infty, \qquad |y| > (3 + \cos x)/2 \\
\end{array} \right\rbrace.\]
As Ansatz for the wave functions, we implemented
\[| \psi \rangle(q,n,k) = \left\lbrace \begin{array}{c}
\frac{1}{2} e^{i \left(\frac{k}{2\pi} + q \right)x} \left(e^{i \frac{n \pi y}{w(x)}} + (-1)^{n+1} e^{-i \frac{n \pi y}{w(x)}}\right), \qquad |y| \leq (3+\cos x)/2 \\
0, \qquad |y| > (3+\cos \pi x)/2 \\
\end{array} \right\rbrace,\]
where $w(x) = \cos (x)+3$ is the width of the channel. The electronic structure for both waveguides was determined numerically as follows. For every value of $m,n$ and $k$, both $|\psi\rangle$ and $\nabla^2 |\psi\rangle$ were calculated on a numerical grid. After normalization, the Hamiltonian matrix elements were calculated as
\[\mathbf{H}_{ij} = \langle \psi_i | \left(V(x,y) |\psi_j\rangle - \frac{1}{2} \nabla^2 | \psi_j \rangle \right),\]
where indices $i$ and $j$ refer to a wave function with a specific set of quantum numbers $q$ (reciprocal-lattice point) and $n$ (transverse wavenumber), and we have set $\hbar = m_e = 1$. Finally, the eigenvalues were obtained for each value of the wave number $k$ by diagonalizing the Hamiltonian matrix.

\begin{figure}[h]
  \centering
  \includegraphics[width=.9\textwidth]{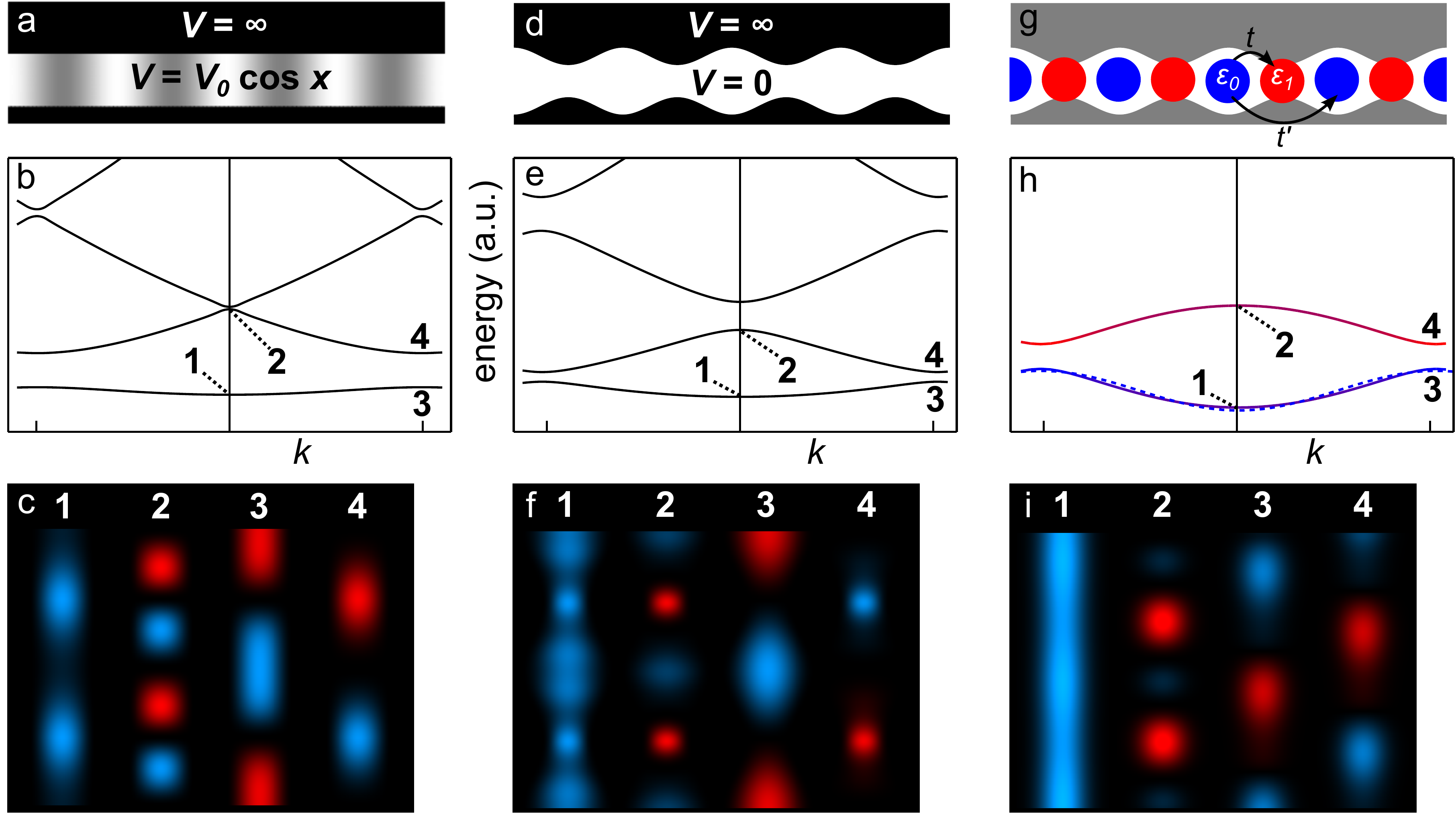}
  \caption{Calculated electronic structure of the potential-modulated waveguide, width-modulated waveguide and two-site tight-binding chain.}
	\label{waveguidedispersions}
\end{figure}

The dispersion for the potential energy-modulated waveguide is shown in Fig.~\ref{waveguidedispersions}. The opening of a band gap is evident at the Brillouin zone boundaries, a very well-known result from the nearly-free electron model. When the crystal orbitals are plotted for this value of $k$ for both bands, the origin of the band gap is shown to arise naturally from orbitals localized either in the potential energy valleys (low energy) or the potential energy hills (high energy). Now, we turn to the result from the width-modulated waveguide. Amazingly, the band structure of the width-modulated waveguide displays exactly the same opening of a band gap at the Brillouin zone boundary. When plotting the crystal orbitals again, we now see that the energy separation arises from electrons localized in the wide regions (low energy) or constrictions (high energy). Therefore, whereas in the first model the electrons are perturbed by the potential-energy landscape, we conclude that in the second model, a similar perturbation arises simply by affecting the geometry of the problem, and therefore the kinetic-energy landscape. \\ \\
Now to finalize the connection to the tight-binding model, we have also approximated the electronic structure of either waveguide in terms of a linear chain of sites. To first-order, these sites may be positioned at the kinetic energy valleys or wide regions (see Fig.~\ref{waveguidedispersions}g). Trivially, this gives a single band described by $E(k) = \epsilon_0 - 2 t' \cos{k}$, where $t'$ is the hopping parameter. Obviously, this model does not allow the description of the localization of the orbitals on the high-energy regions shown (see Fig.~\ref{waveguidedispersions}h, dashed blue line). Therefore, an improved tight-binding model may be constructed by taking both the local energy minimum as well as the interstitial saddle point into account as two separate sites, with the latter one shifted in energy by $\Delta \epsilon = \epsilon_1 - \epsilon_0 > 0$. Now, we obtain two bands with a gap at the Brillouin-zone boundary. These are plotted as the solid curves in Fig.~\ref{waveguidedispersions}h, where the local hue of the band shows the degree of localization on the low-energy sites (blue) and high energy sites (red), respectively. By plotting the crystal orbitals at the same points of the band structure, we obtain a picture similar to the results from the waveguides. As before, the extended model with interstitial high-energy sites functions as an improved description of the electronic structure. \\ \\
We conclude this analysis by noting that the band structure of the Lieb lattice can be set up in a flat energy landscape by confinement effects only. The discretization of the Lieb structure into a tight-binding model is then just as valid, although the interpretation of the model is more subtle since it really relates to the kinetic, rather than potential-energy landscape. Then, to first order, a tight-binding model may be implemented in which only the low-energy regions are assigned as sites. However, a more realistic model, which also describes higher-energy effects, can be obtained by adding sites on the interstitial constrictions, or kinetic saddle points of the lattice. In particular, this inclusion effectively gives new bands at higher energy that show more localization on those high-energy constriction regions.

\subsection{Analytical results from the tight-binding model}

To deepen our understanding of the Lieb lattice and try to obtain a more analytical picture of the band energies, we have analyzed the corresponding tight-binding models in detail. As shown earlier, the Lieb Hamiltonian is given by
\[
\mathbf{H_{Lieb}}=\left(
\begin{array}{ccc}
 \epsilon_1 & -t \gamma_y & -t' \gamma^*_x \gamma_y \\
 -t \gamma^*_y & \epsilon_0 & -t \gamma^*_x \\
 -t' \gamma_x \gamma^*_y & -t \gamma_x & \epsilon_1 \\
\end{array}
\right).
\]
Note that for the absolute square of the variable $\gamma$, the following relation holds
\[|\gamma_x |^2 = \gamma_x \gamma_x^* = (1+e^{i k_x a})(1+e^{-i k_x a}) = 4 \cos^2 \left(\frac{k_x}{2} a \right),\]
and similarly for $\gamma_y$. For convenience, we define the vector $\boldsymbol{\gamma} = (\gamma_x,\gamma_y )$ which has modulus squared
\[||\boldsymbol{\gamma}||^2 = |\gamma_x|^2 + |\gamma_y|^2 = 4\left[\cos^2\left(\frac{k_x}{2} a\right) + \cos^2\left(\frac{k_y}{2} a\right) \right].\]
Note that at any point, the Hamiltonian may be transformed by a similarity transformation:

\[\mathbf{H}\mathbf{c}_i = \mathbf{c}_i E_i\Rightarrow \mathbf{P}\mathbf{H}\mathbf{P}^{-1}\mathbf{P}\mathbf{c}_i = \mathbf{P} \mathbf{c}_i E_i \Rightarrow \mathbf{H}_{sym} \mathbf{c}'_i = \mathbf{c}'_i E_i,\]
\[\mathbf{H}_{sym} = \mathbf{P}\mathbf{H}\mathbf{P}^{-1}\ \ \ \ \ \mathbf{c}'_i = \mathbf{P} \mathbf{c}_i,\]
where the eigenvectors $\mathbf{c}'$ are now in the new basis, consisting of symmetry-adapted linear combinations of site functions. The change of basis is established by choosing the projection operator matrix
\[\mathbf{P} = \left(
\begin{array}{ccc}
 0 & 1 & 0 \\
 \frac{1}{\sqrt{2}} & 0 & \frac{1}{\sqrt{2}} \\
 -\frac{1}{\sqrt{2}} & 0 & \frac{1}{\sqrt{2}} \\
\end{array}
\right),\]
yielding
\[
\mathbf{H}=\left(
\begin{array}{ccc}
 \epsilon_1 & -t \gamma_y & -t' \gamma^*_x \gamma_y \\
 -t \gamma^*_y & \epsilon_0 & -t \gamma^*_x \\
 -t' \gamma_x \gamma^*_y & -t \gamma_x & \epsilon_1 \\
\end{array}
\right)\ \ \ \Leftrightarrow\ \ \ 
\mathbf{H}_{sym} =
\left(
\begin{array}{ccc}
 \epsilon_0  & -t \frac{\gamma_x^*+\gamma_y^*}{\sqrt{2}} & -t \frac{\gamma_x^*-\gamma_y^*}{\sqrt{2}} \\
 -t \frac{\gamma_x+\gamma_y}{\sqrt{2}} & \epsilon_1 - t' \mathrm{Re}\left(\gamma_x \gamma_y^*\right) & + i t' \mathrm{Im}\left(\gamma_x\gamma_y^*\right) \\
-t \frac{\gamma_x-\gamma_y}{\sqrt{2}} & -i t' \mathrm{Im}\left(\gamma_x\gamma_y^*\right) & \epsilon_1 + t' \mathrm{Re}\left(\gamma_x \gamma_y^*\right) \\
\end{array}
\right).
\]
Inspection of these matrix Hamiltonians leads to the conclusion that the lines $\mathbf{M}\rightarrow\mathbf{X}$ and $\mathbf{M}\rightarrow\mathbf{Y}$ are protected from changes in the next-nearest-neighbor hopping. Indeed, here we have either $\gamma_x = 0$ or $\gamma_y = 0$, such that $\gamma_x\gamma_y^* = 0$ and the matrices reduce to
\[
\mathbf{H}_{\mathbf{M}\rightarrow\mathbf{X}/\mathbf{Y}}=\left(
\begin{array}{ccc}
 \epsilon_1 & -t \gamma_y & 0 \\
 -t \gamma^*_y & \epsilon_0 & -t \gamma^*_x \\
 0 & -t \gamma_x & \epsilon_1 \\
\end{array}
\right)=\mathbf{H}(t'=0),\]
\[\mathbf{H}_{sym,\mathbf{M}\rightarrow\mathbf{X}/\mathbf{Y}} =
\left(
\begin{array}{ccc}
 \epsilon_0  & -t \frac{\gamma_x^*+\gamma_y^*}{\sqrt{2}} & -t \frac{\gamma_x^*-\gamma_y^*}{\sqrt{2}} \\
 -t \frac{\gamma_x+\gamma_y}{\sqrt{2}} & \epsilon_1 & 0 \\
-t \frac{\gamma_x-\gamma_y}{\sqrt{2}} & 0 & \epsilon_1 \\
\end{array}
\right)=\mathbf{H}_{sym}(t'=0).
\]
As a result, this Hamiltonian matrix describes both the entire Lieb lattice in absence of next-nearest-neighbor hopping, as well as the more general case where $t'\neq 0$, but then only along the lines connecting the Dirac point with the $\mathbf{X}$ or $\mathbf{Y}$ points. For the latter case, either $k_x = \pi/a$ or $k_y = \pi/a$, which gives $\gamma_x = 0$ or $\gamma_y = 0$. It can be readily seen that then $\mathbf{H}$ becomes block-diagonal, and a trivial eigenvalue of $E_0 = \epsilon_1$ splits off, corresponding to the flat-band energy. Furthermore, at the Dirac point, $\mathbf{H}$ becomes entirely diagonal, resulting in two eigenenergies located at $E = \epsilon_1$ and one eigenenergy at $E = \epsilon_0$. Therefore, it can be seen that the intersection of the flat band with a Dirac cone only exists when the on-site energies are equal: $\epsilon_1 = \epsilon_0$.

In the more general case, the eigenvalues and (normalized) eigenvectors (for either $t'=0$ and $\mathbf{X}/\mathbf{Y}\rightarrow \mathbf{M}$) can be obtained analytically in either basis by diagonalizing the corresponding Hamiltonian. We find

\[E = \left(
\begin{array}{ccc}
E_- \\
E_0 \\
E_+ \\
\end{array}
\right) = \left(
\begin{array}{ccc}
\langle \epsilon \rangle - h_{\mathbf{k}}\left( \frac{\Delta\epsilon}{2} \right) \\
\epsilon_1 \\
\langle \epsilon \rangle + h_{\mathbf{k}}\left( \frac{\Delta\epsilon}{2} \right) \\
\end{array}
\right) \qquad \text{and} \qquad
\mathbf{C} = \left(
\begin{array}{ccc}
\frac{\left(\gamma_y, E_-, \gamma_x \right)}{f_{\mathbf{k}} \left(E_-/2\right)} \\
\frac{\left(\gamma_y,0,-\gamma_x\right)}{f_{\mathbf{k}}(0)} \\
\frac{\left( \gamma_y,E_+, \gamma_x\right)}{f_\mathbf{k} \left(E_+/2 \right)} \\
\end{array}
\right),
\]
where $\Delta \epsilon = \epsilon_1 - \epsilon_0$, $\langle \epsilon \rangle = (\epsilon_0 + \epsilon_1)/2$ and $h_{\mathbf{k}}$ is the hyperbola
\[h_{\mathbf{k}}\left(\frac{\Delta \epsilon}{2}\right) = \sqrt{\left(\frac{\Delta \epsilon}{2}\right)^2 + t^2||\boldsymbol{\gamma}||^2} = \sqrt{\left(\frac{\Delta \epsilon}{2}\right)^2 + 4 t^2 \left[\cos^2 \left(\frac{k_x}{2}a\right) + \cos^2 \left(\frac{k_y}{2}a\right) \right]}.\]
First, we use this result to calculate the energies of the $\mathbf{X}/\mathbf{Y}$-point. The reason is that the density of states shows two sharp peaks at exactly these energies. Evidently, for $k_x = 0$ and $k_y = \pi / a$ or vice versa, one of the cosine functions returns zero whereas the other one gives unity. Therefore, we have
\[||\boldsymbol{\gamma}||^2 = 4t^2\ \ \ \Rightarrow\ \ \ E_{\pm,\mathbf{X}/\mathbf{Y}} = \langle \epsilon \rangle \pm \sqrt{ \left( \frac{\Delta \epsilon}{2} \right)^2 + (2t)^2},\]
which in the case of equal on-site energies reduces to $E_{\pm,\mathbf{X}/\mathbf{Y}} = \epsilon_0 \pm 2t$.
Secondly, this result shows that a Dirac cone emerges at $\mathbf{M}$ only if $\Delta \epsilon = 0$, in which case an expansion of $\mathbf{k}$ around $\mathbf{M}$ leads to
\[E_{\pm} = \epsilon_0\pm 2t \sqrt{\cos^2 \left(\frac{k_x}{2}a\right) + \cos^2 \left(\frac{k_y}{2}a\right)} \approx \epsilon_0\pm t||\mathbf{k}||a.\]
Here, we use the fact that the dispersion is smooth, and therefore the effect of $t'$ converges to zero towards the $\mathbf{M}$-point, even along the lines of \emph{k}-space where it may not be cancelled out. The Fermi velocity in the Dirac cone can be calculated as
\[v_{F} = \frac{1}{\hbar} \frac{\partial E_{\pm}}{\partial ||\mathbf{k}||} = \frac{t a}{\hbar}.\]
However, if the on-site energies of the sites are not equivalent, the Dirac cone splits up into a hyperbola, where the flat band stays attached to either the top band in case of higher edge-site energies, or the bottom band in case of higher corner-site energies.

From the experimentally obtained spectra, we could not observe a band gap, which suggests $\Delta \epsilon \approx 0$ within the limits of experimental broadening. However, as noted in the manuscript, an additional peak in the spectra on the edge sites at 90 mV suggests that the ``flat band'' is not exactly flat but may be curled up, generating a new van Hove singularity in the density of states at the energy corresponding to its maximum. Therefore, it makes sense to investigate the effect of next-nearest-neighbor hopping. We started the investigation by calculating the eigenenergies along the line $\mathbf{\Gamma}\rightarrow\mathbf{M}$. This line connects the Dirac point with the $\mathbf{\Gamma}$-point, where the flat band is observed to `curl up' (see Fig.~\ref{squareLiebsuperLieb}). Along this line, we can substitute $\gamma_x = \gamma_y = \gamma = 1 + e^{i k_{(x/y)} a}$, which gives
\[
\mathbf{H}_{\mathbf{\Gamma}\rightarrow\mathbf{M}}=\left(
\begin{array}{ccc}
 \epsilon_1 & -t \gamma & -2 t' |\gamma|^2 \\
 -t \gamma^* & \epsilon_0 & -t \gamma^* \\
 -2 t' |\gamma|^2 & -t \gamma & \epsilon_1 \\
\end{array}
\right)\ \ \ \Leftrightarrow\ \ \ 
\mathbf{H}_{sym, \mathbf{\Gamma}\rightarrow\mathbf{M}} =
\left(
\begin{array}{ccc}
 \epsilon_0 & -\sqrt{2} \gamma t & 0 \\
 -\sqrt{2} \gamma^* t & \epsilon_1 - t' |\gamma|^2 & 0 \\
 0 & 0 & \epsilon_1 + t' |\gamma|^2 \\
\end{array}
\right).
\]
In this case, the symmetrized Hamiltonian has become block-diagonal, which means that the last row of $\mathbf{P}$ - a SALC describing a pure edge-localized state - coincides with an eigenvector of the Hamiltonian
\[\mathbf{p}_3 = \mathbf{c}_0 = \left(\frac{1}{\sqrt{2}},0,-\frac{1}{\sqrt{2}}\right)\ \ \ \Rightarrow\ \ \ E_0 = \epsilon_1 + t' |\gamma|^2,\]
where we may substitute
\[|\gamma|^2 = 4 \cos^2\left(\frac{k_{(x/y)} a}{2}\right) = 2 \left[ \cos^2\left(\frac{k_x a}{2}\right) + \cos^2\left(\frac{k_y a}{2}\right) \right] = \frac{||\boldsymbol{\gamma}||^2}{2}\]
since along $\mathbf{\Gamma}\rightarrow \mathbf{M}$, $k_x = k_y$. Hence, the ``flat band'' is curved up towards $\mathbf{\Gamma}$, and achieves its maximum value of $E_{0,\mathbf{\Gamma}} = \epsilon_1 + 4 t'$ there. Visual inspection confirms that this is the maximum energy of the band, which means that it corresponds to a Van Hove-singularity in the density of states. As a result, the second peak in the experimentally obtained spectra on the edge sites may be thought to originate from this feature.

The remaining energy eigenvalues of the top and bottom bands can be obtained by diagonalizing the remaining block,
\[\mathbf{H}_{sym\pm, \mathbf{\Gamma}\rightarrow\mathbf{M}} =
\left(
\begin{array}{cc}
 \epsilon_0 & -\sqrt{2} \gamma t \\
 -\sqrt{2} \gamma^* t & \epsilon_1 - t' |\gamma|^2 \\
\end{array}
\right),
\]
\[E_\pm = \langle \epsilon \rangle - t' \frac{||\boldsymbol{\gamma}||^2}{4} \pm \sqrt{ \left( \frac{\Delta \epsilon}{2} - t' \frac{||\boldsymbol{\gamma}||^2}{4} \right)^2 + t^2 ||\boldsymbol{\gamma}||^2 }.\]
When expanding around $\mathbf{M}$, to first order, only terms linear in $||\mathbf{k}||$ remain. In this case, the dispersion reduces to the form where $t'$ vanishes, yielding again the Dirac cone (or hyperbola in case of different on-site energies). At the $\mathbf{\Gamma}$ point, however, the top and bottom bands are perturbed,
\[E_{\pm,\mathbf{\Gamma}}(t'=\Delta\epsilon =0) = \epsilon_0 \pm 2\sqrt{2} t\ \ \ \Rightarrow\ \ \ E_{\pm,\mathbf{\Gamma}}(\Delta\epsilon =0,t'\neq 0) = \epsilon_0 - 2t' \pm \sqrt{ \left(2\sqrt{2} t\right)^2 + (2 t')^2}.\]
As a result, $t'$ widens the bandwidth but also shifts the average of the $\mathbf{\Gamma}$-point energies with respect to the Dirac point. Since $t'$ is assumed to be a negative parameter (bonding next-nearest neighbor interaction), its effect is to effectively compress the top band, lowering it towards the Dirac point energy, whereas the bottom band moves to even lower energy.

\begin{figure}[h]
  \centering
  \includegraphics[width=.8\textwidth]{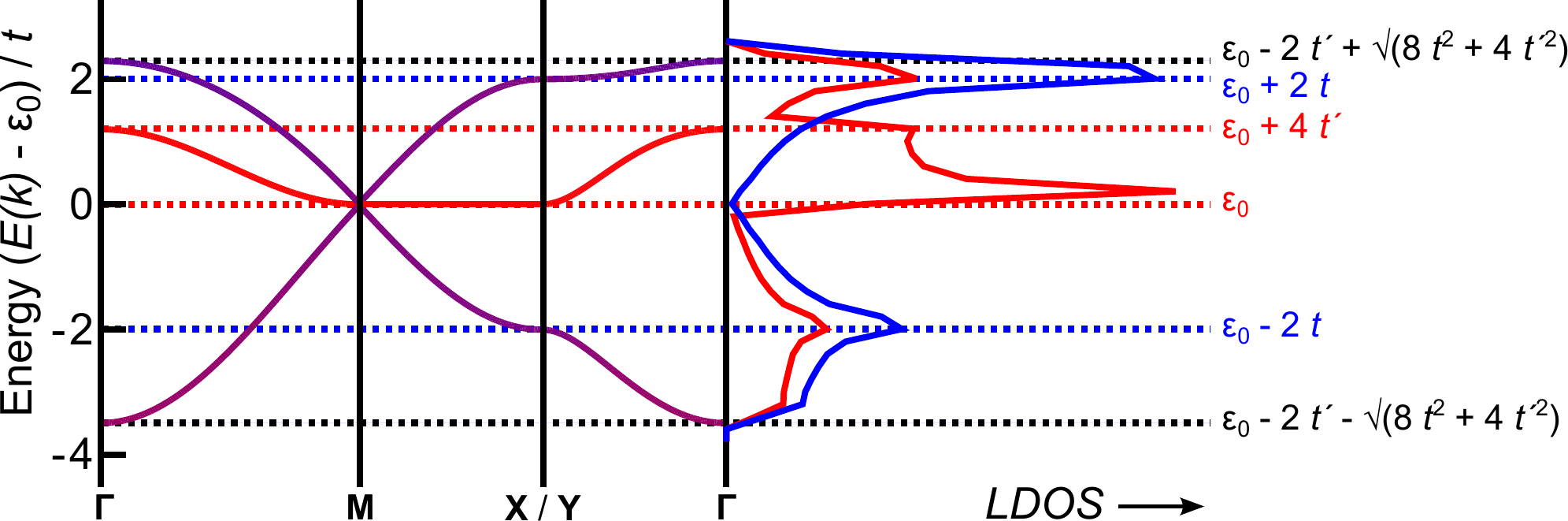}
  \caption{Tight-binding dispersion of the Lieb lattice for $\Delta \epsilon = 0$ and $t' \neq 0$, with all features in the density of states labelled analytically.}
	\label{dispersion_analytical}
\end{figure}

With the information above, we labelled all peak features in the density of states analytically. The obtained knowledge was utilized to find the best parameters to describe the experimental data. The dispersion and local density of states with analytical peak labels is shown in Fig.~\ref{dispersion_analytical}.

\subsection{Finite-size effects}

\begin{figure}[h]
  \centering
  \includegraphics[width=.7\textwidth]{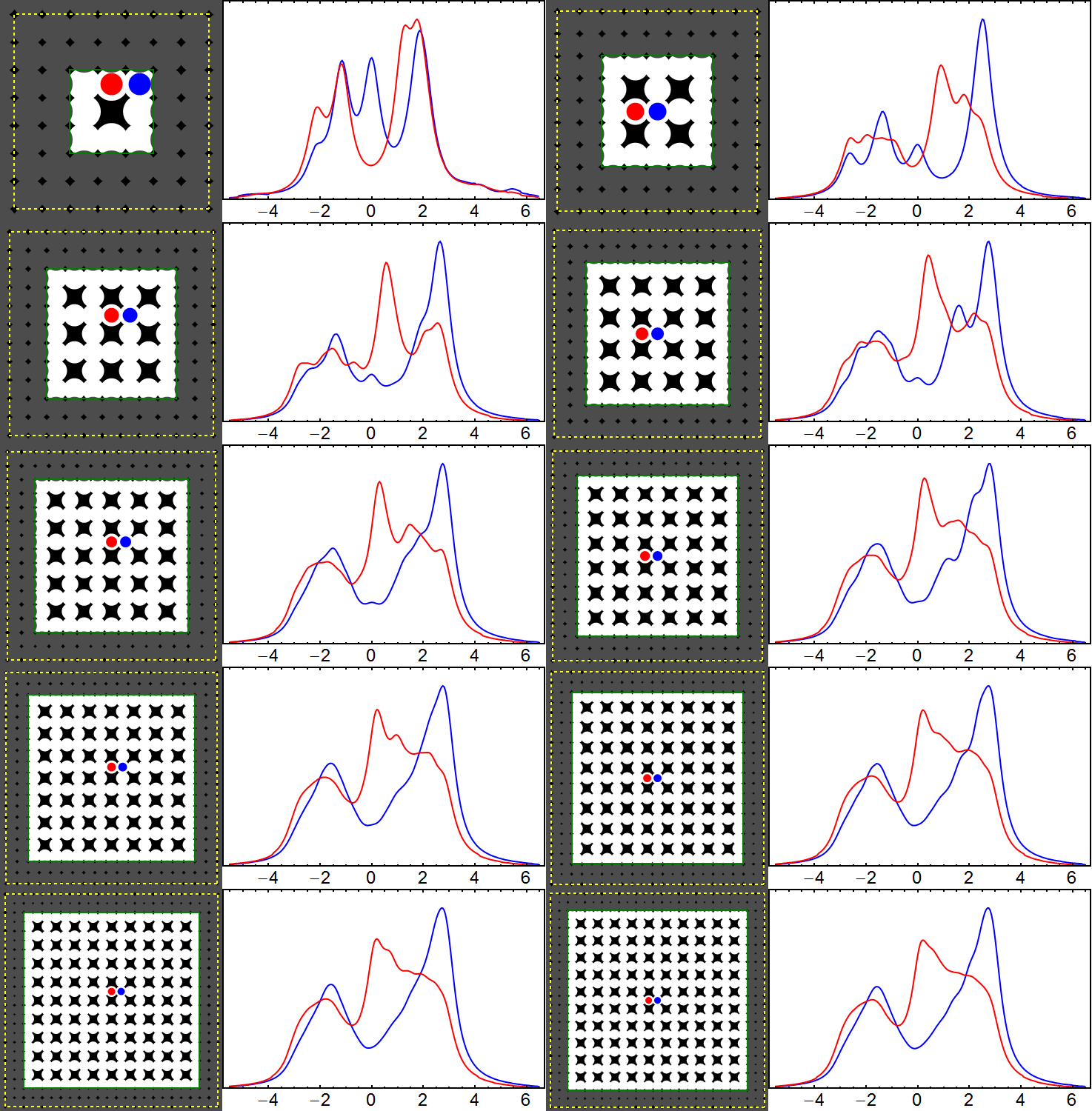}
  \caption{Tight-binding simulations for Lieb lattices with 1~x~1 up to 10~x~10 unit cells. Hard-wall boundary conditions were used. The 5~x~5 configuration corresponds to the experimentally realized geometry.}
	\label{finitesizes_hardwalls}
\end{figure}

In addition to the periodic-lattice calculations, the tight-binding method was used to model the experimentally realized finite-size lattices. Importantly, we wanted to find out to what extent finite-size effects play a role in lattices of these small dimensions.
To this end, the local density of states was calculated on corner and edge sites for a range of different lattice sizes. Fig.~\ref{finitesizes_hardwalls} shows the results for lattice sizes of 1 by 1 up to 10 by 10. Here, the LDOS spectra are displayed for a corner site (blue curve, indicated by blue dot on the lattice) and an adjacent edge site (red curve, indicated by red dot on the lattice), as closely as possible to the middle of the lattice. These calculations employed hard-wall boundary conditions, where $\Psi = 0$ at and beyond the lattice boundary. As can be seen from Fig.~\ref{finitesizes_hardwalls}, for lattices of size 5 by 5 and 6 by 6, the edge site spectra contain a peak at the energy corresponding to the flat band at $\mathbf{\Gamma}$, corresponding to the experimental observations (\emph{cf.} Fig.~2b). For smaller sizes, the shape of the spectra is still not ``stable'', whereas for larger lattice sizes the shape converges and smoothens out.

In the experimental lattices, there is obviously no hard wall between the electronic states in the lattice and the unconfined 2DEG around it. As a crude model of the surrounding metal, we therefore took ``2DEG'' sites around the Lieb lattice into account by means of an effective square-lattice model. To account for the larger spectral range and lower Fermi level of the 2DEG, we used a lower on-site energy ($\epsilon_{2DEG} = \epsilon_0 - 0.4 t$) and higher bandwidth through increased coupling strength ($t_{2DEG} = 2 t_{lattice}$, $t'_{2DEG} = 2t'_{lattice}$, $s_{2DEG} = 2 s_{lattice}$). Hopping between 2DEG sites and lattice sites was described by the lattice hopping parameters.

\begin{figure}[h]
  \centering
  \includegraphics[width=.7\textwidth]{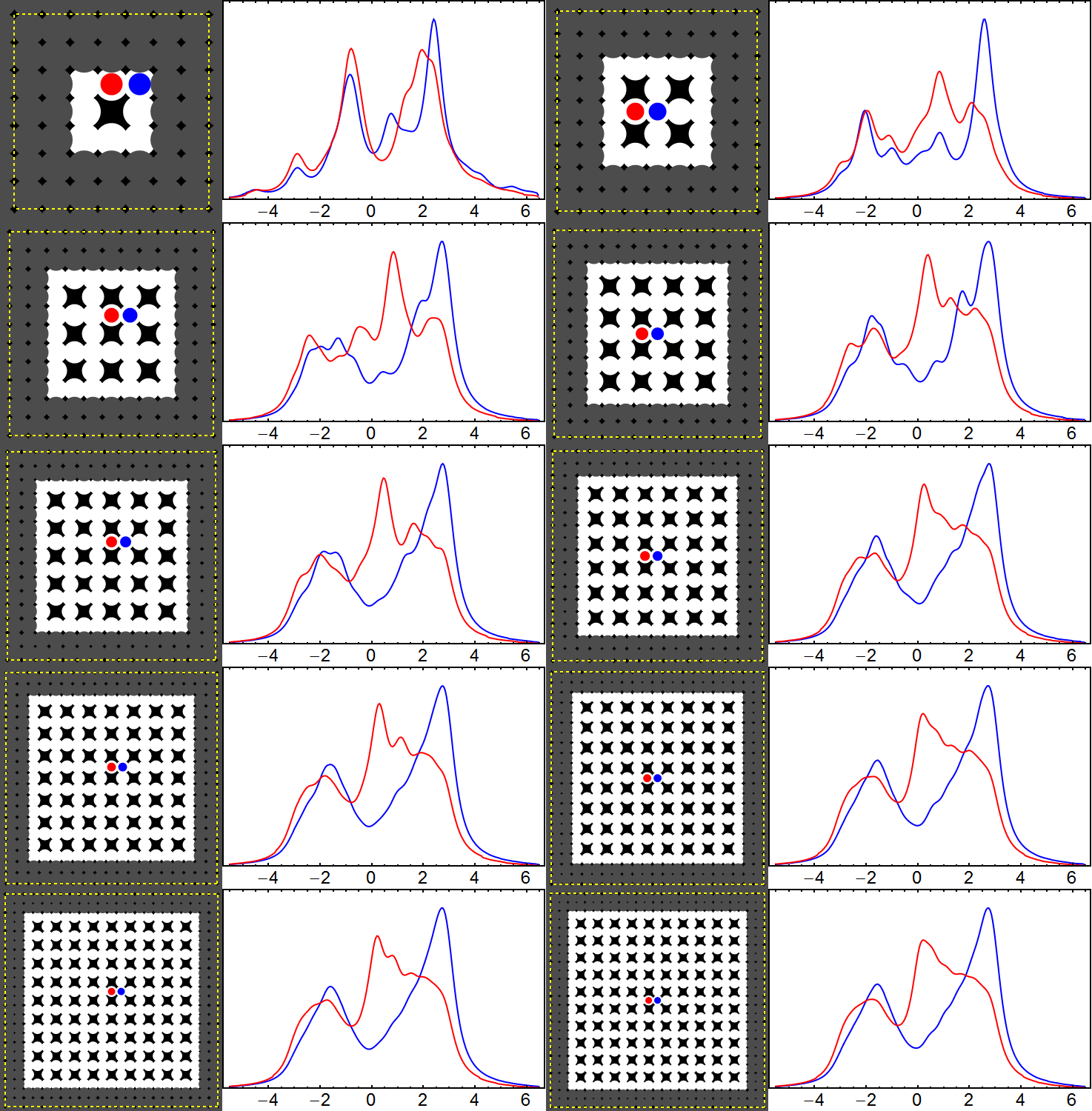}
  \caption{Tight-binding simulations for Lieb lattices with 1~x~1 up to 10~x~10 unit cells. Periodic boundary conditions were used, \emph{i.e.} no walls, in contrast to the hard-wall boundary conditions in Fig.~\ref{finitesizes_hardwalls}. The 5~x~5 configuration corresponds to the experimentally realized geometry.}
	\label{finitesizes_nowalls}
\end{figure}

In order to investigate the influence of this hopping, we again calculated the LDOS for lattices of increasing size. In contrast to the hard-wall boundary calculations, these calculations have therefore used open-wall boundary conditions. The results are shown in Fig.~\ref{finitesizes_nowalls}. When comparing the LDOS spectra of Fig.~\ref{finitesizes_nowalls} with Fig.~\ref{finitesizes_hardwalls}, we note a significant difference for the smallest lattices, but a very quick convergence to the hard-wall boundary results. Already for lattices of size 4 by 4, it is difficult to notice any difference between the results with hard walls and open walls. We therefore conclude that the electronic structure in the lattice is really a property of the lattice and is not significantly perturbed by the surrounding 2DEG. \\
The tight-binding calculations on finite-size lattices use periodic boundary conditions on the 2DEG ``boundary sites''. Since the unit cell is quite large, and we found that the electronic structure inside the lattice is only weakly affected by the surrounding 2DEG, we only sampled the Brillouin zone of the entire geometry at the $\Gamma$-point. Although this limits the number of Bloch waves taken into account inside the 2DEG, the effect on the wave functions localized inside the lattice is minimal. 

\newpage
\section{Experimental results}

\subsection{Differential conductance spectra}
The normalized differential conductance spectra shown in the paper were obtained by dividing the differential conductance spectra over the Lieb and square lattice by an average of spectra acquired on the clean Cu(111) surface, analogously to the normalization in Ref. \cite{Gomes2012}. Division of the spectra on the Lieb and square lattice by the copper spectrum cancels the contribution of the density of states of the tip and the slope of the Cu(111) spectrum, yielding the normalized spectra presented in Fig.~2b and 2e. 

\begin{figure}[hb]
	\centering
	\includegraphics[width=0.5\textwidth]{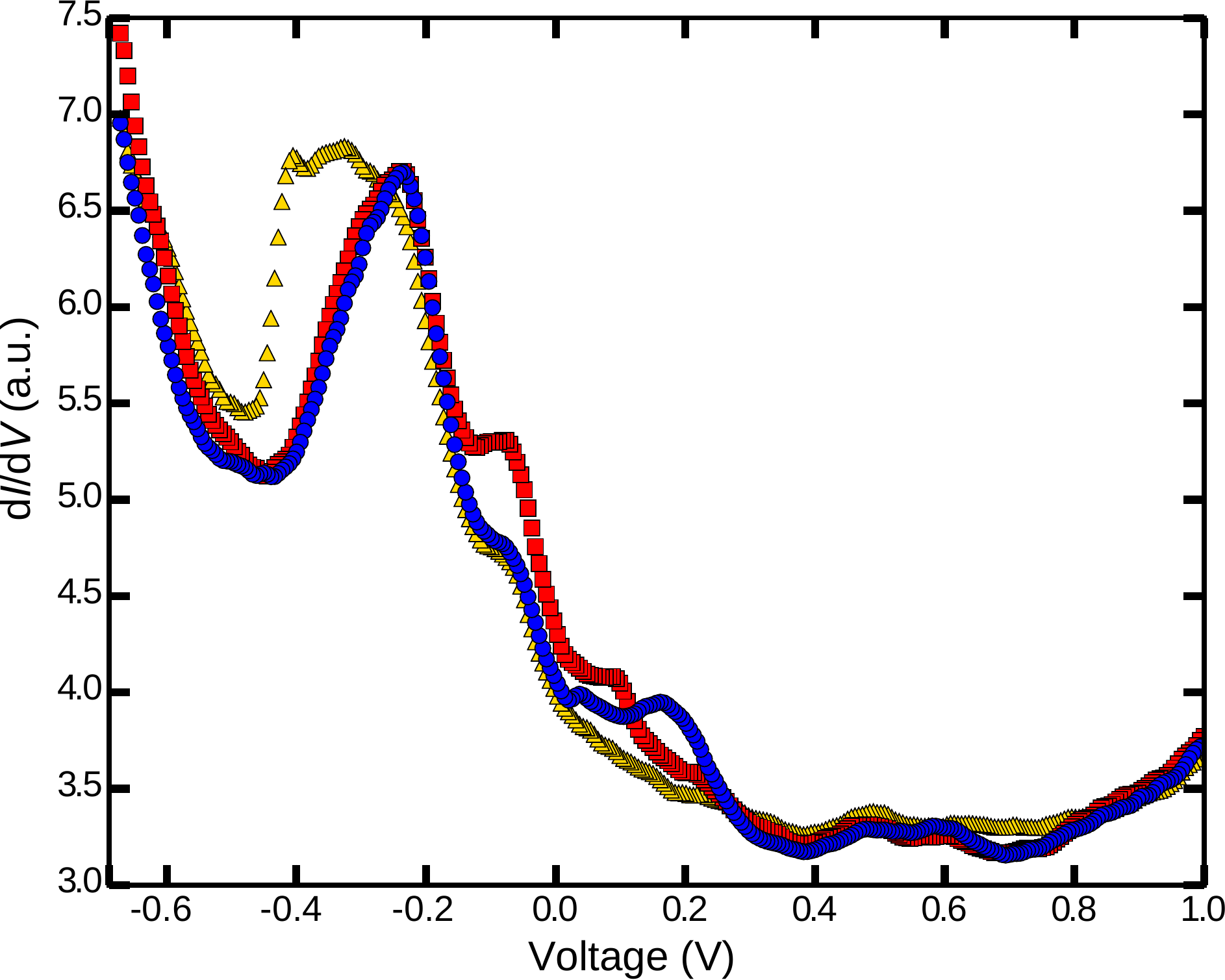}
	\caption{\textbf{Differential conductance spectra $\vert$} d$I$/d$V$ spectrum on a corner (blue) and edge (red) site in the $5\times 5$ Lieb lattice compared with the corresponding d$I$/d$V$ spectrum on bare Cu(111) (yellow).} 
	\label{rawspectrum}
\end{figure}

In Fig.~\ref{rawspectrum}, the spectra on the corner (blue) and edge sites (red) of the Lieb lattice from Fig.~2b are presented without normalization. The corresponding average copper spectrum is shown in yellow. The peaks characteristic for the Lieb lattice can be observed on top of the spectrum on Cu(111). Various different tips ($>25$), characterized by differently shaped Cu(111) spectra, were used to corroborate the features arising from the Lieb lattice. For each tip, the average copper spectrum is an average of between 3 and 100 spectra randomly taken on different positions on the clean Cu(111). 

\subsection{Influence unit-cell size}

\begin{figure}[!h]
	\centering
	\includegraphics[width=0.8\textwidth]{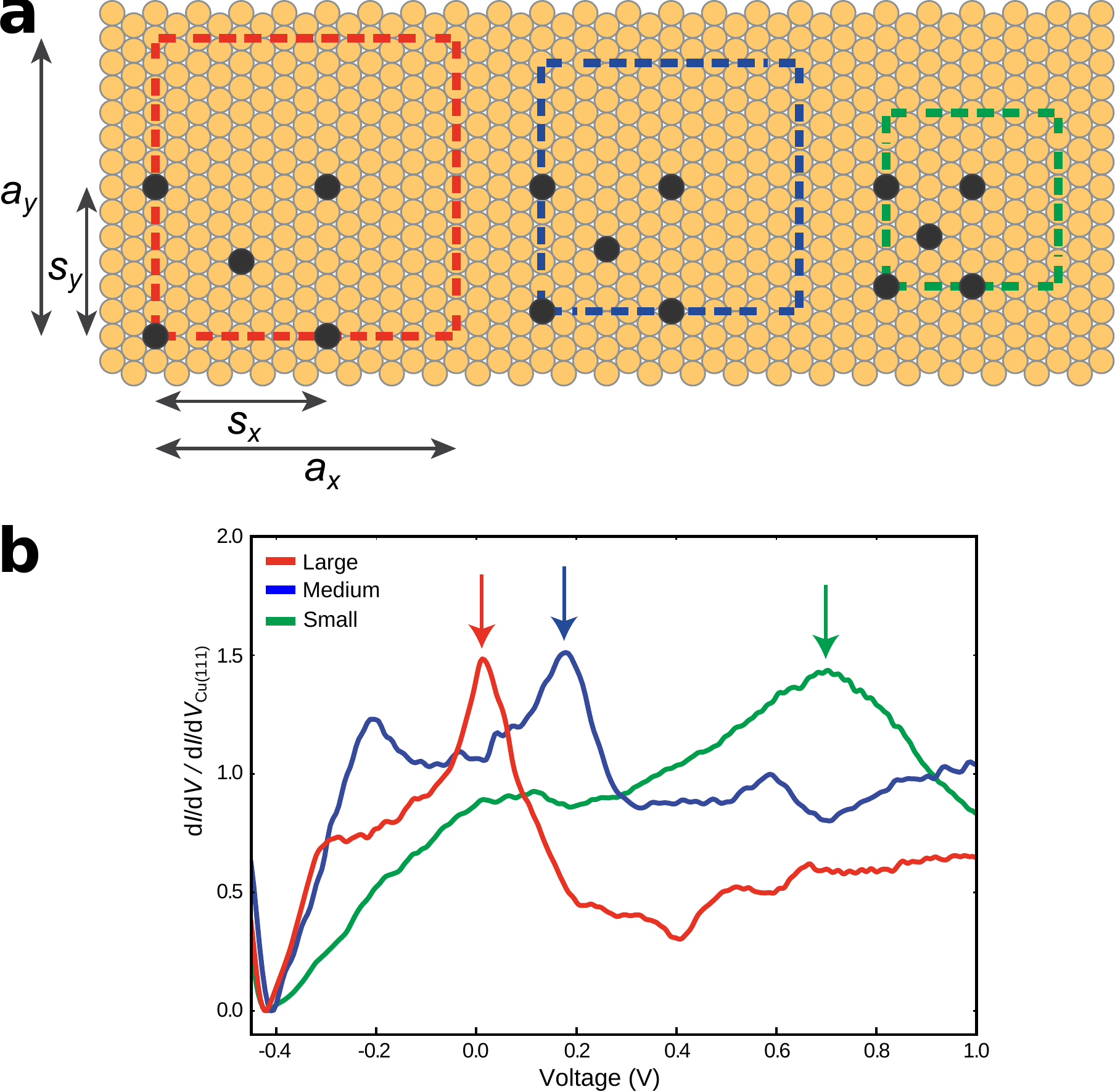}
	\caption{\textbf{Unit cell geometries.}  \textbf{a,} The CO molecules on the top-sites of the Cu(111) surface (gray) compose a $3.10\,$nm $\times$ $3.07\,$nm (red), $2.66\,$ $\times$ $2.56\,$nm (blue), and $1.77\,$nm $\times$ $1.79\,$nm (green) unit cell. The $2.66\,$nm by $2.56\,$nm unit cell (blue) represents the chosen configuration for the measured Lieb lattices. \textbf{b,} Normalized differential conductance spectra acquired above corner sites for the unit cells in \textbf{a}. The arrows indicate the peak assigned to the highest-energy band of the Lieb lattice and show a shift towards higher energies for smaller unit-cell sizes.} 
	\label{DesignSize_SI}
\end{figure}

The size of the unit cell has a large influence on the position of the peaks in the differential conductance spectra \cite{Gomes2012}. Since there is no significant charge transfer between the CO molecule and the Cu(111) surface, the electron density of the Cu(111) surface is largely unaffected by the number of adsorbed CO molecules \cite{Ropo2014}. If we increase the unit cell, the number of surface-state electrons in the unit cell increases. This shifts the energy bands of the designed lattice with respect to the Fermi level of the underlying Cu(111). The unit-cell size can thus be used to tune the lattice into an $n-$doped (large unit cell) or $p-$doped (small unit cell) structure, as demonstrated for the graphene geometry by Gomes \emph{et al.} \cite{Gomes2012}. \\ \\
In Fig.~\ref{DesignSize_SI}a, a schematic representation of several geometries of the unit cell of the Lieb lattice is shown. Lattices with these configurations were built and characterized using scanning tunneling spectroscopy. The differential conductance spectra in Fig.~\ref{DesignSize_SI}b confirm the shift of the peaks with respect to the Fermi level as a function of the unit-cell size: the smaller the unit cell, the further the peaks are shifted to positive voltages. We obtain an $n-$doped structure for a large unit cell of $3.10\,$nm by $3.07\,$nm (shown in red) and a $p-$doped structure for a small unit cell of $1.77\,$nm by $1.79\,$nm (shown in green). For a unit-cell size of $2.66\,$nm by $2.56\,$nm, we obtain a close to neutral structure, which is chosen for further measurements. Importantly, not only this unit cell is square to a good approximation ($a_x / a_y \approx 1.04$), but also the CO molecules are positioned in an approximately square, face-centered configuration ($s_x / s_y \approx 1.04$). Thus, the assigned sites in the Lieb lattice preserve their symmetry and the two edge sites in the unit cell retain their geometric equivalence with a deviation below 4$\%$.

\clearpage

\end{document}